\documentclass[sn-mathphys]{sn-jnl}

\jyear{2021}%

\theoremstyle{thmstyleone}%
\theoremstyle{thmstyletwo}%

\theoremstyle{thmstylethree}%

\begin{document}

\title[Optimizing data selection protocols for tumor model calibration]{Utilizing gradient approximations to optimize data selection protocols for tumor growth model calibration}


\author*[1]{\fnm{Allison L.} \sur{Lewis}}\email{lewisall@lafayette.edu}

\author[1]{\fnm{Kathleen M.} \sur{Storey}}

\author[2]{\fnm{Heyrim} \sur{Cho}}

\author[1]{\fnm{Anna C.} \sur{Zittle}}

\affil[1]{\orgdiv{Department of Mathematics}, \orgname{Lafayette College}, \orgaddress{\street{730 High St}, \city{Easton}, \postcode{18042}, \state{PA}, \country{USA}}}

\affil[2]{\orgdiv{Department of Mathematics}, \orgname{University of California, Riverside}, \orgaddress{\street{900 University Ave}, \city{Riverside}, \postcode{92521}, \state{CA}, \country{USA}}}

\abstract{The use of mathematical models to make predictions about tumor growth and response to treatment has become increasingly more prevalent in the clinical setting. The level of complexity within these models ranges broadly, and the calibration of more complex models correspondingly requires more detailed clinical data. This raises questions about how much data should be collected and when, in order to minimize the total amount of data used and the time until a  model can be calibrated accurately. To address these questions, we propose a Bayesian information-theoretic procedure, using a gradient-based score function to determine the optimal data collection times for model calibration.  The novel score function introduced in this work eliminates the need for a weight parameter used in a previous study's score function, while still yielding accurate and efficient model calibration using even fewer scans on a sample set of synthetic data, simulating tumors of varying levels of radiosensitivity. We also conduct a robust analysis of the calibration accuracy and certainty, using both error and uncertainty metrics. Unlike the error analysis of the previous study, the inclusion of uncertainty analysis in this work---as a means for deciding when the algorithm can be terminated---provides a more realistic option for clinical decision-making, since it does not rely on data that will be collected later in time.}




\maketitle

\section{Introduction}

In recent decades, mathematical modeling has frequently been used to advance our understanding of tumor evolution \cite{Altrock2015,Lavi2012,Swierniak,Byrne2010,Rockne2019}. Modeling of cancer can be performed from the complex, highly-refined cellular level to a more ``macro" level view, where we assume that the tumor acts as a mass of homogeneous tissue.  Estimating the parameter values of such models requires detailed data, which may take many forms \cite{Chambers2019,Bibault}. The models can then be used to make predictions about the evolution of the tumor and its response to various treatment modalities, including radiotherapy, chemotherapy, immunotherapy, and viral therapy, among others. Recent technological advances have made it possible to collect a wide variety of data describing tumors, from the molecular level to the tissue level.  Collecting data at multiple time points can aid in the calibration of mathematical models, which can be tailored to incorporate the available data. However, some data collection can be prohibitively expensive or invasive; this raises questions about how much data is needed to make accurate clinical predictions using mathematical models, and when this data should be collected.   

In the age of personalized medicine, clinicians are turning to individualized treatment protocols, each tailored to the unique patient. Mathematical modeling can play a significant role here; given data from an individual tumor, we can calibrate a model and determine patient-specific parameter values which may give insight into the efficacy of the proposed treatment regimen for that individual. However, it is important that we bridge the gap between the idealized math modeling framework and the clinical constraints. While highly complex models can be insightful as far as determining the underlying mechanisms of the tumor and predicting how different cell populations might interact, at the clinical level, we are very constrained in the level of detail that might be inferred from the available data. The question then is: can an inherently simplistic model calibrated solely from a very small budget of crude data (i.e. estimated tumor volume from an MRI scan) still yield useful information regarding predicted response to treatment? 

Because data collection in a clinical oncology setting is both expensive and potentially invasive for the patient, clinicians are constrained to a very sparse budget of measurements.  Practically speaking, a clinician might collect a tumor volume scan at diagnosis, a second one at the start of treatment, and then neglect to measure again until the treatment period has ended. With such sparse data, it can be difficult to construct a model with any sort of predictive power; the amount of uncertainty in such a model will be prohibitive. Thus, we wish to investigate how one might get the most ``bang for their buck" for a specified data collection budget. If we are restricted to $n$ data points, at what points should we collect them?  What time periods during the treatment regimen are most informative, in terms of reducing the uncertainty of the model parameters? 

In \cite{ChoJCM}, an algorithmic approach to determining an optimal selection of scans for model calibration was proposed by the authors. This approach relied on a Bayesian information-theoretic sequential experimental design framework, in which each data point was chosen in turn by maximizing a given score function, whereupon the model parameters were re-calibrated to give an updated model trajectory. The score function utilized in \cite{ChoJCM} was proposed as a means of adapting the pre-existing sequential design framework to handle time-series data, as opposed to other studies \cite{Lewis,Terejanu} which dealt solely with non-temporal data (i.e. spatial design conditions).  In addition to trying to maximize the reduction in parameter uncertainty through the choice of a highly informative data point, we also sought to penalize the algorithm for skipping too many data points, since the temporal data framework does not allow for those points to be subsequently collected at a later date. This penalization step, at the time, relied upon a penalization parameter $k$, which we varied over the interval $[0,1]$ in an attempt to optimize the efficiency and accuracy of the model calibration. The previous study tested this algorithm on three sets of synthetic data of varying radiation response types, and concluded that the optimal $k$ value varies depending on the strength of patient response to the radiotherapy treatment.  For instance, in scenarios where the tumor was highly sensitive to radiation, the model calibration procedure benefited most from the use of a $k$ value near or at 1. Scenarios with data that was less responsive tended to favor $k$ values in the low-to-middle spectrum, i.e. $k=0$ to $k=0.3$.

Although this framework was demonstrated to be effective in determining which scans to select for model calibration, the previous study did have several weaknesses. Most notably, the reliance of the choice of parameter value $k$ upon the shape of the patient data was constrictive; an optimal $k$ value could not be determined until the general shape of the data could be assessed, which required at least several data points. In a highly restrictive scan budget scenario---i.e., in the clinical scenarios we are attempting to mimic---this means that an optimal $k$ value realistically cannot be determined in time to have a positive impact on the algorithm efficiency. Thus, finding a way to eliminate dependence upon the penalization parameter value is a focus of this work; in particular, we propose using information gathered about gradient approximations to adapt the weighting of the penalization term as the algorithm progresses, in place of using a static parameter $k$. 

Additionally, we conduct an analysis of this new gradient-based score function with mean-square error, as was used in \cite{ChoJCM}, and we supplement this with uncertainty-based analysis, using credible intervals constructed by propagating parameter posterior distributions through the model to assess the level of certainty in the resulting model trajectory. The uncertainty analysis relies solely on the data that has been collected up to the current day, so it provides a more practical assessment of confidence in the model predictions for use in a clinical setting.

We begin in Section \ref{sec:models} by describing the low-fidelity ordinary differential equation model that we'll use throughout the investigation to illustrate the algorithm.  Additionally, we give a brief background about the source of the synthetic data used---obtained from a cellular automaton model---and describe how our virtual patient cohort was developed. Section \ref{sec:methodology} outlines the algorithm development, including the necessary background in Bayesian parameter estimation and sequential design, and the formulation of the new score function. Our metrics for model assessment are discussed in Section \ref{sec:assessment}.  Section \ref{sec:results} first compares the results from the new score function to those obtained using the score function from the previous study \cite{ChoJCM}. We conclude in this section that the penalization parameter $k$ can now be discarded, and we present the remainder of the model calibration results for three spheroids of varying radiotherapy sensitivity.  We conclude Section \ref{sec:results} with an analysis of how the model uncertainty is affected by measurement noise. Section \ref{sec:discussion} summarizes the findings of the investigation and discusses their implications.

\section{Low-Fidelity Model and Synthetic Data} \label{sec:models}

In this section, we present the low-fidelity differential equation model that we use throughout this investigation to demonstrate the algorithm.  Because this is a proof-of-concept investigation that requires comparing errors and uncertainties across different collections of scans and taking data measurements as often as once per day, the study is ideally suited for the use of synthetic data.  Thus, we also introduce a more complex stochastic cellular automaton model that will act as the high-fidelity model to generate synthetic data in place of experimental data; we consider the data generated from the CA model to be our ``truth" and calibrate the low-fidelity model to fit the high-fidelity data. 

\subsection{Low-Fidelity Differential Equation Model}  \label{sec:lofimodel}

The low-fidelity model that we use for calibration is an ODE model that describes the total tumor volume over time. The model describes the time evolution of the total tumor volume, $V(t)$, using a logistic growth model with growth rate $\lambda$ and carrying capacity $K$:

\begin{equation}
    \frac{dV}{dt} = \underbrace{\lambda V \left ( 1 - \frac{V}{K} \right )}_{\mbox{logistic growth}} - 
    \underbrace{\eta V.}_{\mbox{natural cell death}}
    \label{eqn1comp_dimensional}
\end{equation}
 We note that in this form, Equation (\ref{eqn1comp_dimensional}) is not structurally identifiable \cite{ChoSpringer}, since there exist infinitely many pairs ($\lambda$,$\eta$) which yield the same net growth rate $\lambda-\eta$. 
Thus, we re-parameterize Equation (\ref{eqn1comp_dimensional}) to obtain the parametrically-identifiable form 

\begin{equation}
    \frac{dV}{dt} = AV\left(1-\frac{B}{A}V\right),
    \label{eqn:eqn1comp}
\end{equation}
where $A = \lambda-\eta$ and $B = \frac{\lambda}{K}$. Moving forward, any reference to the low-fidelity model refers to the re-parameterized form, Equation (\ref{eqn:eqn1comp}).

In this simple one-compartment model, we assume that any dead or necrotic cells from sustained oxygen or nutrient deprivation are removed from the tumor instantaneously; that is, we view the tumor as a homogeneous mass of proliferating, viable cells. 

We incorporate the radiotherapy (RT) treatment protocol using the linear-quadratic model \cite{Hall1994, Enderling} to account for the effects of RT. In this model, the fraction of cells that survive exposure to a single administered dose $d$ of RT is given by
\begin{equation}
\mbox{Survival fraction},\ \  SF = e^{-\alpha d-\beta d^2},
\label{eq:SF}
\end{equation}
where $\alpha$ and $\beta$ represent tissue-specific radiosensitivity parameters (for single-strand and double-strand DNA breaks, respectively), and $d$ is the radiotherapy dosage. We incorporate a typical radiotherapy regimen for solid tumors, with daily doses of 2 Gy administered Monday through Friday for six consecutive weeks. We note that the linear-quadratic model is a reasonable choice for fast-growing tumors \cite{Perez-Garcia2015}. We assume that the irradiated cell fraction is removed immediately from the tumor volume, similarly to the instantaneous removal of necrotic cells in our model. We reformulate the low-fidelity model with radiotherapy, under these assumptions, as:

\begin{align}
   \left\{
\begin{array}{ll}
      \frac{dV}{dt} = A V \left ( 1 - \frac{B}{A}V \right ),  & \mbox{for} \; t_i^+ < t < t_{i+1}^-,\\ \ \\
      V(t_i^{+})=
\exp(-\alpha d-\beta d^2) \: V(t_{i}^{-}), &\\
\end{array} 
\right. 
\end{align}
where $t_i$ (for $i=1, 2, \ldots, n_R$) denote the times at which an RT dose is delivered, and $V(t_i^{\pm})$
denote the tumor volume just before and after radiotherapy is administered. Previous work (\cite{ChoSpringer}) has illustrated that the full parameter set $[A, B, \alpha,\beta]$ is unidentifiable in the sense that multiple sets of parameters may yield the same model output. As such, we fix the $\alpha/\beta$ ratio to be 1.5 and estimate $\beta$ only, in addition to $[A,B]$ \cite{brenner1999}.  We acknowledge that the assumption of exponential treatment response may overestimate radiation sensitivity, and in the future we plan to compare the results of our methodology using different radiotherapy models \cite{Prokopiou2015,Rockne2010,Poleszczuk2018,Sunassee2019}.

\subsection{High Fidelity Data} \label{sec:camodel}

For this proof-of-concept work, in place of experimental data we generate high fidelity data by using a hybrid cellular automaton (CA) model, adapted from the models developed in \cite{Paczkowski2021,ChoSpringer,ChoJCM}. 
We use this model to generate a series of synthetic spheroids that differ in their response to radiosensitivity. In the model, cells are arranged on a discrete lattice representing a two-dimensional square cross-section of size 0.36$\times$0.36 cm$^2$ through a three-dimensional cancer spheroid \emph{in vitro}. Notable features of the model include a heterogeneous cancer population and stochastic cell cycle, coupled with spatially heterogeneous oxygen levels modeled by a partial differential equation.

 Each automaton $\textbf{x}=(x,y)$ can be occupied at time $t$ by a tumor cell in one of three states---proliferating, $\mathcal{P}$, quiescent, $\mathcal{Q}$, or necrotic, $\mathcal{N}$---or can be unoccupied and denoted as empty, $\mathcal{E}$. We assume oxygen is the single growth-rate-limiting nutrient, so the cell state is determined by the spatially heterogeneous oxygen level, modeled as a reaction diffusion equation. See \cite{Paczkowski2021,ChoJCM} for a detailed description of the oxygen model. 
We note that all lattice sites have an associated oxygen level, $c$, which determines the state of the occupying cell using thresholds $c_N$ and $c_Q$: if $c > c_Q$ then the cells proliferate, if $c_N < c < c_Q$ then the cells transition to quiescent cells with a smaller oxygen consumption rate, and if $c \leq c_N$ then the cells are considered necrotic. 

If a cell becomes necrotic due to low oxygen concentration, the necrotic cells are lysed at rate $p_{NR}$.
Lysis involves removing the necrotic cell and then shifting inward a chain of cells starting from the boundary of the spheroid to fill in the removed cell's site. 
  Additionally, cells in the model divide more slowly when they have a large number of neighbors, mimicking the regulatory process known as contact inhibition of proliferation. After cell division, chains of cells are shifted outward, simulating cell-cell adhesion. The details of these model features are described in \cite{Paczkowski2021,ChoJCM}.

The baseline parameter values that are used to generate data using the CA model are shown in the Appendix \ref{app:params}, and detail about the parameters that we vary to generate distinct synthetic spheroids is provided in Section \ref{sec:virtual_cohort}. These parameter values are estimated using experimental data from the prostate cancer cell line, PC3, in \cite{Paczkowski2021}. Note that the parameter values are listed with volumetric units; we convert the units to a two-dimensional cross sectional area by assuming that the three-dimensional tumor takes on an ellipsoidal shape.

Radiotherapy in the CA model is implemented using the linear-quadratic model detailed in equation \eqref{eq:SF}. The probability of survival for all proliferating cells in the CA model is identical to the survival fraction in the low-fidelity model. In the CA model, the quiescent cells are $\frac{2}{3}$ times as likely as the proliferating to be irradiated, in order to reflect the lower sensitivity to radiation-induced DNA damage for quiescent cells, in comparison to proliferating cells.

\subsection{Virtual Spheroid Cohort}\label{sec:virtual_cohort}


We generated a virtual cohort of 27 tumor spheroids using the CA model described in Section \ref{sec:camodel}, for calibration testing. In order to generate spheroids with a range of  responses to radiotherapy, we varied the mean cell cycle time, $\bar{\tau}_{cycle}$ (in hours), the radiosensitivity parameter $\alpha$, and the ratio $\alpha/\beta$. We generated one virtual spheroid with each combination of parameter values listed in the ranges below, while fixing all other parameter values at the values listed in Table \ref{table:CA_pars}.
\begin{align*}
&\bar{\tau}_{cycle} \in [15, \, 22, \, 30], \\
&\alpha \in [0.014, \, 0.5, \, 0.14], \\
&\alpha/\beta \in [1, \, 1.5, \, 2].
\end{align*}

\begin{figure}[htbp]
\centerline{  
         \includegraphics[width=0.33\linewidth]{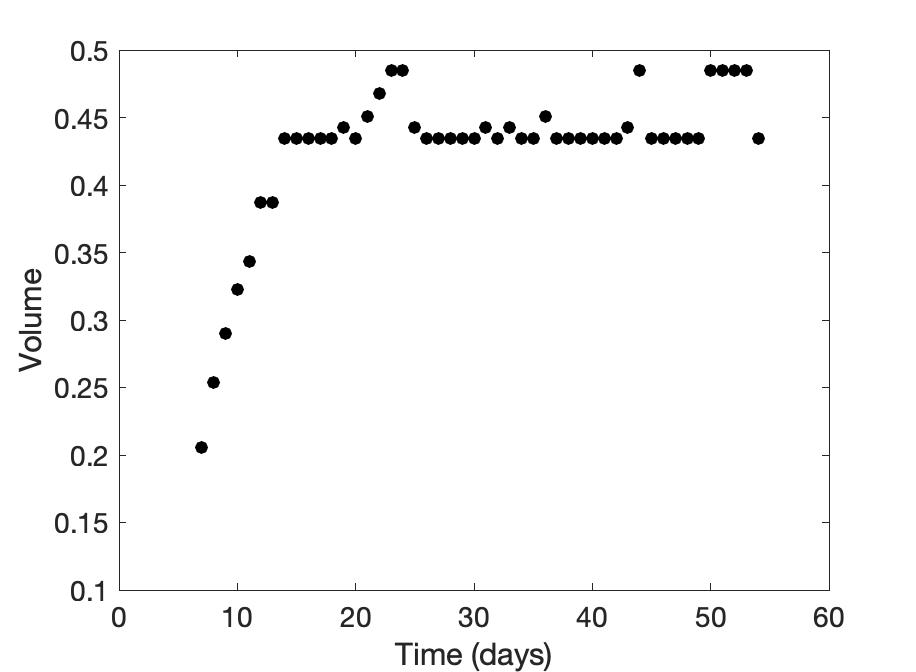}
         \includegraphics[width=0.33\linewidth]{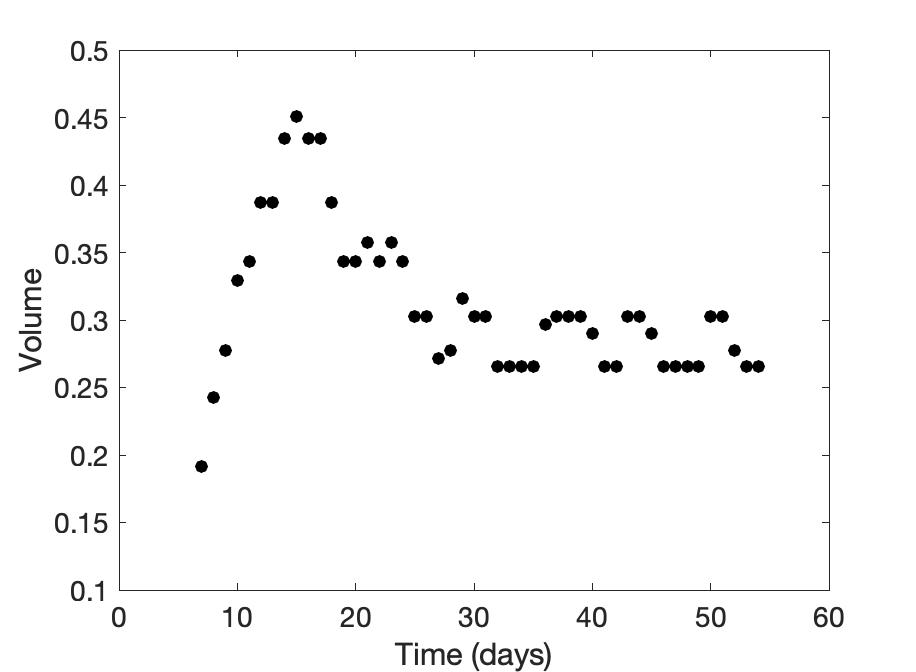}
         \includegraphics[width=0.33\linewidth]{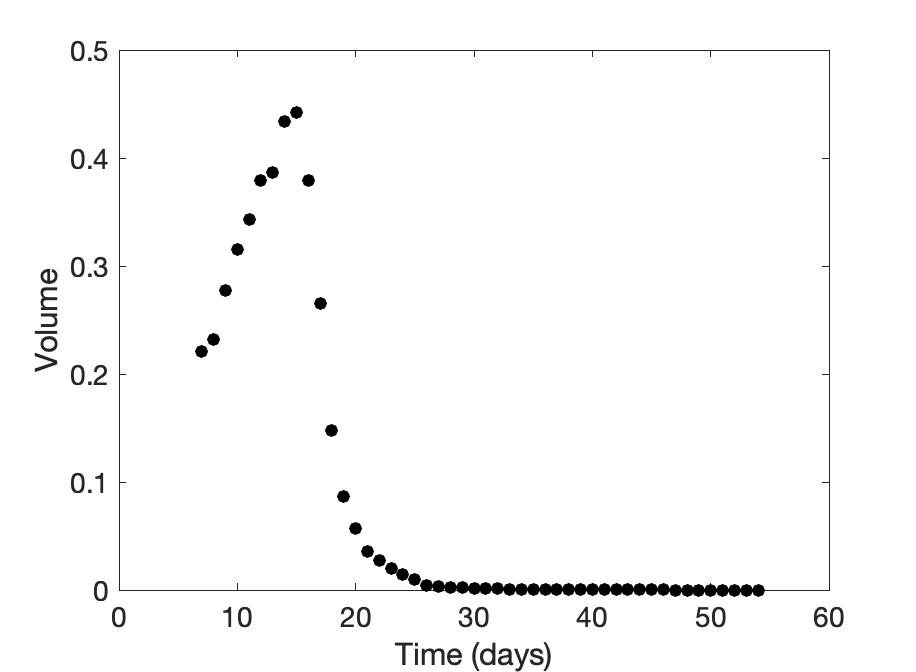}
}
    \caption{High fidelity data of tumor volume trajectory for our three virtual patients: non-responder (left), medium responder (middle), and strong responder (right). The data is generated by the hybrid CA model described in Section \ref{sec:camodel} using parameters given in Section \ref{sec:virtual_cohort}.  }
    \label{fig:patients}
\end{figure}

Next, by visually inspecting the simulation results, we separated the 27 virtual spheroids into three categories: non-responders, medium responders, and strong responders. We observed similar patterns among the spheroids in each category, with respect to the quality of fits and to the timing of and number of scans chosen using the original score function and the new gradient-based score function. For simplicity we chose one representative from each category to present in our results section, each of which reflects roughly the average calibration behavior across the simulations in its category. Our chosen representative non-responder was generated using the parameter values $\bar{\tau}_{cycle}=22$, $\alpha=0.014$, and $\alpha/\beta=1$. Our chosen representative  medium responder was generated using the parameter values $\bar{\tau}_{cycle}=22$, $\alpha=0.05$, and $\alpha/\beta=1.5$, and our chosen representative strong responder was generated using the parameter values $\bar{\tau}_{cycle}=22$, $\alpha=0.14$, and $\alpha/\beta=1$. 
The synthetic tumor volume data of the three virtual patients are shown in Figure \ref{fig:patients}.  


\section{Methodology for Optimal Data Collection} \label{sec:methodology}

In this section we introduce the methodology used to determine optimal time points at which to collect data in order to best inform our model parameters.  We begin with a brief discussion of Bayesian parameter estimation techniques---used to re-calibrate our parameter set upon each data point acquisition---then briefly outline the sequential experimental design setup.  As these two concepts formed the basis of the previous study, we direct the reader to \cite{ChoJCM} for further details. We finish this section with the introduction of our new gradient-based score function for balancing the choice of an informative data point with the need to gather data early during the treatment period, so that treatment protocols can be altered in the event that the model trajectory predicts an undesirable outcome.  

\subsection{Bayesian Parameter Estimation} \label{sec:bayesian}

Throughout this investigation, we rely on Bayesian methods for parameter estimation and mutual information construction.  Because the Bayesian perspective assumes that parameters are random variables with associated densities that can be repeatedly updated to reflect new information from data acquisition, these methods are ideally suited for our sequential design framework.  Additionally, the construction of posterior densities that reflect uncertainty in the parameter values allows for the propagation of these densities through the model to quantify the resulting uncertainty in the output. Thus, Bayesian methods are ideal for uncertainty quantification, as will be further discussed in Section \ref{sec:uncanalysis}. 

In the Bayesian framework, we encode previously known information about each parameter value (i.e. physical bounds or a hypothesized distribution) into a prior distribution. We then update our prior distributions based on information gained from additional data to construct posterior distributions for each parameter.  For prior distribution $p(\theta)$ and likelihood function $p(D_n \mid \theta)$ (which quantifies the likelihood of observing data set $D_n$ given parameter values $\theta$), Bayes' Rule gives the resulting posterior distribution as $$p(\theta \mid D_n) = \frac{p(D_n \mid \theta)p(\theta)}{p(D_n)}.$$ 

In practice, posterior distributions are constructed through the use of Metropolis algorithms, which sample parameter candidates from throughout the parameter space, compute the likelihood function for each candidate, and compose a chain of accepted candidates from which the posterior distribution is formed. In this study, we specifically use the Delayed Rejection Adaptive Metropolis algorithm, which incorporates (a) a mechanism for updating (or adapting) the covariance matrix as information is gained, and (b) the proposal of a secondary parameter candidate in the event that the first is rejected, thereby mitigating the issue of chain stagnation.  Further details about the DRAM algorithm and Metropolis algorithms in general can be found in \cite{Haario, Smith}.

Within this investigation, the DRAM algorithm is used to re-calibrate our parameter set $[A, B, \beta]$ from the model outlined in Section \ref{sec:lofimodel} after the selection of each additional data point for collection.  Data points are chosen from a sequential design framework, as outlined in the next section. 

\subsection{Sequential Experimental Design} \label{sec:seqdesign}

Recall that our overarching goal is to accurately calibrate our model using as little data as possible.  As such, we need a way to determine which potential data points will be most informative for our parameter set; that is, given a choice of potential scan day options, which collection of scans will maximize the reduction in parameter uncertainty?  

To answer this question, we utilize a sequential experimental design framework, in which data points are acquired one-by-one and parameter estimates are updated between each data acquisition. For a current data set $D_{n-1} = \{\tilde d_1, \tilde d_2, \dots, \tilde d_{n-1}\}$ and a set of possible design conditions $\Xi$, we select the design condition $\xi_n\in \Xi$ that will maximize the reduction in uncertainty of the model parameters $\theta$ when $\tilde d_n$---the data point resulting from collecting experimental (or synthetic) data at condition $\xi_n$---is added to the existing data set.  We can quantify the information contribution of design $\xi_n$ upon parameter set $\theta$ by computing the mutual information, 
\begin{eqnarray}
     &I&(\theta; d_n \mid D_{n-1},\xi_n) = \nonumber \\
     \nonumber \\
     & & \int_{\mathcal{D}} \int_{\Omega} p(\theta, d_n \mid D_{n-1},\xi_n) \log \frac{p(\theta, d_n \mid D_{n-1},\xi_n)}{p(\theta \mid D_{n-1})p(d_n \mid D_{n-1},\xi_n)} d\theta d d_n
\label{eq:mi_eqn}
\end{eqnarray}
where $d_n$ represents the predicted value of $\tilde d_n$ using our model, 
$\mathcal{D}$ is the full set of all unknown future observations, 
and $\Omega$ describes the multivariate parameter space.  The mutual information provides a measure of parameter uncertainty reduction; a larger MI value indicates potential for a greater acquisition of knowledge about the parameter value than a small MI value.  For more details on the derivation of mutual information and computational methods used to estimate it, we point the reader to \cite{ChoJCM, Lewis, Terejanu, Kraskov}.

In a standard non-temporal sequential design framework that utilizes MI as a metric, we would compute the MI for each of the potential design conditions, then choose the condition which maximizes the MI as our next condition for experimental or synthetic evaluation.  After evaluation of this data point, the parameter set is re-calibrated and the computation of MI begins anew. The algorithm can be terminated when either (a) a user-defined threshold for model accuracy or uncertainty is achieved, or (b) a pre-defined budget of scans is exhausted.

For a scenario such as this investigation, in which design conditions represent temporal points at which data can be collected, we require an adaptation to the standard methodology.  Because collecting data at time $t_n$ precludes the collection of data for all times $t_i$ with $i<n$, we must account for the potential loss of information from skipped data points. In our previous study \cite{ChoJCM}, we proposed an adaptation to the MI framework using a score function that would reward a user for choosing a point with a large MI but simultaneously penalize them for skipping other points. In the following section, we amend this proposal to employ information about the approximate gradient in the score function, allowing for better optimization of our algorithm without the need for additional parameters.

\subsection{Gradient-Based Score Function} \label{sec:newscorefxn}

In the previous section, we discussed the goal of calibrating our model using as little data as possible. However, our overall goal is actually two-fold; while we want to use as little data as possible, we also prefer that the collected data be obtained from early during the treatment period. This early calibration allows for changes to be made to the treatment regimen if the model prediction reveals that the current regimen will not be effective.

This two-part goal introduces a trade-off to consider: we want to reward the choice of design conditions that yield a large mutual information (i.e.~those that maximize the reduction in parameter uncertainty), but we also want to penalize design conditions that require us to skip far ahead in time for data collection.  In pursuit of this goal, we outline the following metric for choosing the optimal data point for collection.

Suppose that the current data consists of data set $D_r = \{\tilde d_1,\tilde d_2,\dots, \tilde d_{r}\}$, where $\tilde d_{r} = \tilde d(t_{r})$ is the most recently appended data point. Among all possible future data points, $\tilde d_{r+1}, \tilde d_{r+2}, \dots \tilde d_n$, we wish to determine which point will yield the most information about our model parameters. We define the relative mutual information at step $r$, $$R(i,r) = \frac{I(\theta; d(t_i) \mid D_r)}{I_r^*},$$ to be a normalized metric for quantifying the mutual information between low-fidelity model prediction $d(t_i)$ and parameter set $\theta = [A,B,\beta]$ for design condition $i$; note that $I_r^*$ represents the maximum mutual information seen across all potential design conditions. The relative mutual information by itself gives us a quantity to be maximized.  We can also summarize the potential information loss from skipping points $r+1$ through $i-1$ using the ratio $$\frac{\sum\limits_{j=r+1}^{i-1} R(j,r)}{\sum\limits_{l=r+1}^{n_T}R(l,r)},$$ which adds up the relative mutual information of all skipped points and divides by the total relative mutual information across all possible data points.  The combination of these two terms, where the second term is weighted by a penalty parameter $k$, was presented in \cite{ChoJCM} as

\begin{equation}
S_k^{\text{old}}(i,r) = R(i,r)-k\left(\frac{\sum\limits_{j=r+1}^{i-1} R(j,r)}{\sum\limits_{l=r+1}^{n_T}R(l,r)}\right).
\label{eq:oldScore}
\end{equation}

This old score function was shown to successfully find an appropriate scanning protocol for distinct patients with different radiotherapy response patterns \cite{ChoJCM}. 

However, although we were able to find recommended scanning protocols using \eqref{eq:oldScore}, the scanning schedule and calibration accuracy was highly reliant on the penalty weight parameter $k$. For highly responsive patients, larger values of $k$ close to one gave more accurate results, while less responsive patient data was calibrated more accurately with small values of $k$ close to zero. That is, calibrating a model to data with rapidly changing dynamics benefits from the collection of more data than is necessary for calibration to a data set with a more gradual change. This need for more frequent data collection could be achieved by choosing larger values of weight $k$ to assign a larger penalty for skipping days. 

Inspired by this pattern, we now aim to modify the previous score function to have less dependency on the penalty weight parameter. To quantify the patients' response to radiotherapy, we compute the rate of change of the  tumor volume and incorporate that into the score function. In particular, the rate of change is used to scale the penalty ratio to mimic the pattern of more responsive patients requiring larger values of $k$. The rate of change is computed by the change in volume over the last two data points. We use a first order finite difference scheme to approximate the gradient of the tumor volume measurement, but use relative change in volume to keep it appropriately scaled. The approximate gradient of the relative tumor volume is thus defined as  
$$\frac{\tilde d_r-\tilde d_{r-1}}{\tilde d_{r-1} (t_r-t_{r-1})}.$$ 
Then, by multiplying the absolute value of the relative gradient by the penalty summation term, we define our new score function as 
\begin{equation}
S_k(i,r) = R(i,r)-k\cdot\bigg \lvert \frac{\tilde d_r-\tilde d_{r-1}}{\tilde d_{r-1}(t_r-t_{r-1})}\bigg \rvert \left(\frac{\sum\limits_{j=r+1}^{i-1} R(j,r)}{\sum\limits_{l=r+1}^{n_T}R(l,r)}\right), 
\label{eq:NewScore} 
\end{equation}
assuming $\tilde d_{r-1} \neq 0$.
We use the absolute value of the gradient since our intuition behind scaling $k$ is to give more penalty when data rapidly changes, which includes both increasing and decreasing dynamics. Computationally, if $\tilde d_{r-1}$ is zero, we set $S_k(i,r) = R(i,r)$; that is, we ignore the temporal penalty term, since unchanging data suggests that we can skip later in time without missing vital information. In Section \ref{sec:scorecomp}, we analyze whether the inclusion of the weight parameter $k$ is still necessary in this new formulation.

\section{Analyzing Model Accuracy and Uncertainty} \label{sec:assessment}
Section \ref{sec:methodology} outlined a procedure for choosing optimal design conditions at which to collect data in a sequential manner.  But when should we terminate the algorithm?  The user has two options.  If there are no constraints on the data collection budget, the user might define a goal that they wish to meet in terms of model accuracy or reduction of uncertainty; this might take the form of a user-defined error or uncertainty threshold.  Once this goal is attained, the user may terminate the algorithm.  In a more likely scenario, there are significant constraints on the data collection budget due to expensive or invasive scanning procedures that will force termination of the algorithm. For this scenario, the user must determine whether an adequate reduction in uncertainty or error has been achieved.  This analysis will assist the user in deciding whether the resulting model is reliable enough to be used for decision-making at the clinical level.  

Though the previous study relied solely on error analysis to determine the predictive power of the final model, here we expand our model assessment to include an uncertainty analysis component. This is a more suitable assessment method in practice, since a user can measure the level of uncertainty in the model at any point using only the data collected so far, but cannot measure the full error in the model until after data collection has ceased. 

\subsection{Error Analysis} \label{sec:erroranalysis}

Our previous study \cite{ChoJCM} relied solely on error analysis as a means of determining the number of scans required to achieve model accuracy.  We conduct that analysis here again, and compare how the goals of achieving model accuracy (i.e. error reduction) and model certainty (i.e. uncertainty reduction) align.

To assess model error, we calculate the mean-square-error between the low-fidelity model and the high-fidelity synthetic data for all possible scan choices, given by $$\text{MSE} = \frac{1}{n-1}\sum\limits_{i=1}^{n} (y_i - f(x_i; \theta))^2,$$ where $y_i$ represents the high-fidelity synthetic data measurement on day $i$ and $f(x_i; \theta)$ represents the model prediction at day $i$ given parameter set $\theta$. For this particular investigation, we include data for days 7-56, representing one week prior to the treatment initiation and all six weeks of treatment.  We use this metric to demonstrate how the low-fidelity model trajectory converges toward the ``truth" data as the scan number increases, thus showing the improvement of model accuracy with increasing scan number.

The drawback to using this metric for model assessment is that in practice, one would not have access to all of these high-fidelity data evaluations for computation. Given only the data points about which the user is actually aware, it is difficult to assess whether the model parameters have converged to the values that will create the idealized model fit across the entire data regime.  Using only the selected scans, a final error could potentially be informative, but this is in essence a ``hindsight" analysis; we cannot compute this error until all data have been collected. Thus, we supplement our error analysis in this investigation with an uncertainty analysis, and investigate how either one might be used to assess convergence towards the ideal model fit.

\subsection{Uncertainty Analysis} \label{sec:uncanalysis}

The use of Bayesian methods for parameter estimation provides an ideal setting for performing uncertainty quantification. The posterior distributions for each of the parameter values can be propagated through the model to simulate the full array of resulting trajectories that might arise. In essence, one can directly observe the uncertainty in the model output that arises from the uncertainty in the parameter inputs.  In this study, we construct 95\% credible intervals for the model output by feeding the parameter chains from the Metropolis algorithm through the model, then plotting the middle 95\% of trajectories to accompany the chosen model fit (the fit that utilizes the set of parameter values which is found to maximize the likelihood function).  

As a metric for quantifying the amount of uncertainty in our model predictions, we estimate the area of the credible interval at each data acquisition step.  As the uncertainty in the model parameters is reduced with each new added data point, this manifests as a tighter credible interval about the fitted model trajectory; we can observe how the area of the interval trends generally downward with each new data collection.

The major benefit to conducting uncertainty analysis---as opposed to error analysis---is that it can be considered ``foresight analysis".  Given only the data that we have already collected, we can measure the uncertainty in the model trajectory for future times and assess whether this uncertainty has been reduced to an acceptable level to allow for decision-making based on the model. However, it should be noted that just because the model uncertainty has been reduced to an acceptable level does not guarantee that the model fit to future data will be decent. We recommend a two-fold approach: waiting for the model uncertainty to be reduced while also checking that the model trajectory has stabilized across the previous few data additions; seeing the trajectory change drastically with each added point suggests that the model may require additional data in order to settle upon a best fit. 

\section{Results} \label{sec:results}

We compare the calibration results using our proposed gradient-based score function defined in Section \ref{sec:newscorefxn} with the results using the original score function we defined in \cite{ChoJCM}. In our score function comparison, we consider three sample spheroids, chosen to represent a strong responder, medium responder, and non-responder to radiotherapy, as shown in Figure \ref{fig:patients}. For each sample spheroid, we compare the chosen scans using each score function and the error and uncertainty for a set of scan budgets, while varying the score function weighting parameter $k$. We initialize our algorithm with four pre-treatment data points on days 7-10, so that the parameter set $[A,B,\beta]$ can be initially calibrated. We note that the required pre-treatment data could be reduced to two points by using a universal growth parameter $A$ and an average radiotherapy response rate $\beta$.

\subsection{Score Function Comparison} \label{sec:scorecomp}

\begin{figure}[htbp]
\centerline{  
         \includegraphics[width=0.49\linewidth]{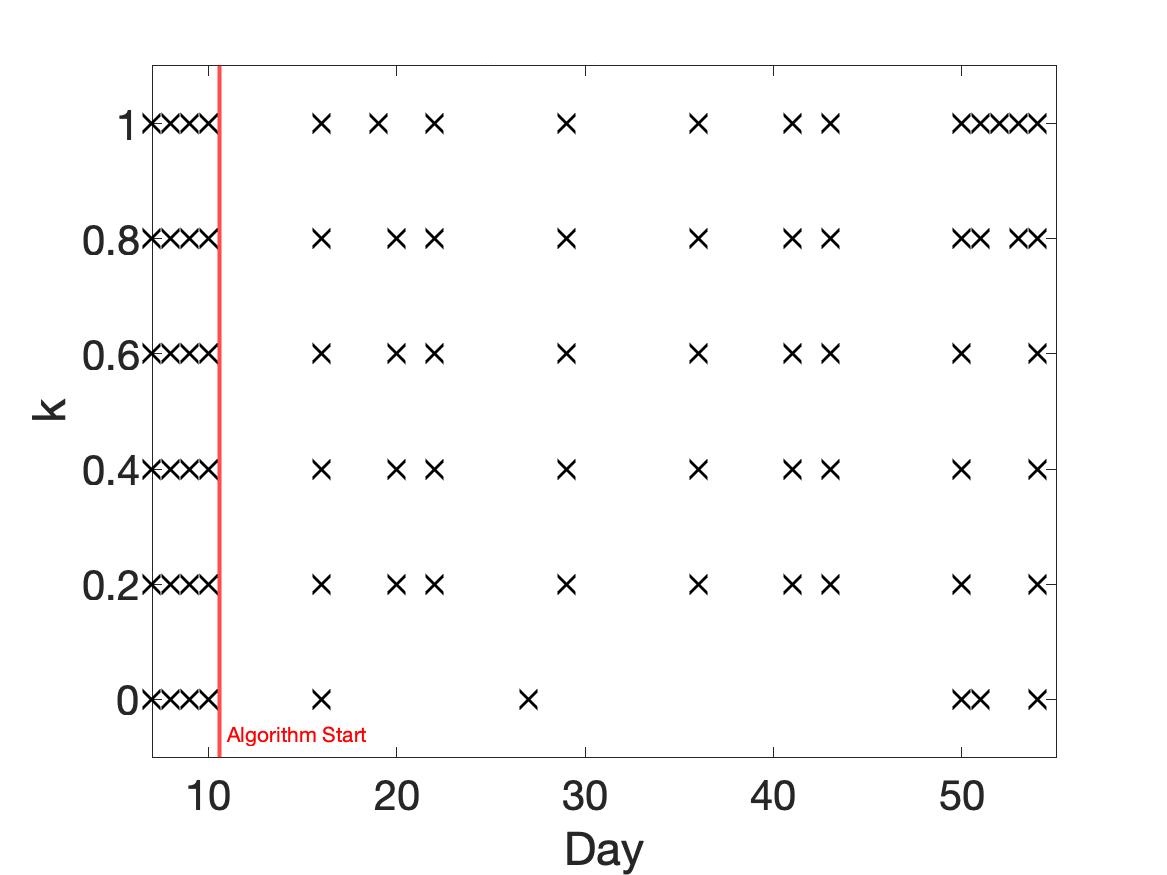}
         \includegraphics[width=0.49\linewidth]{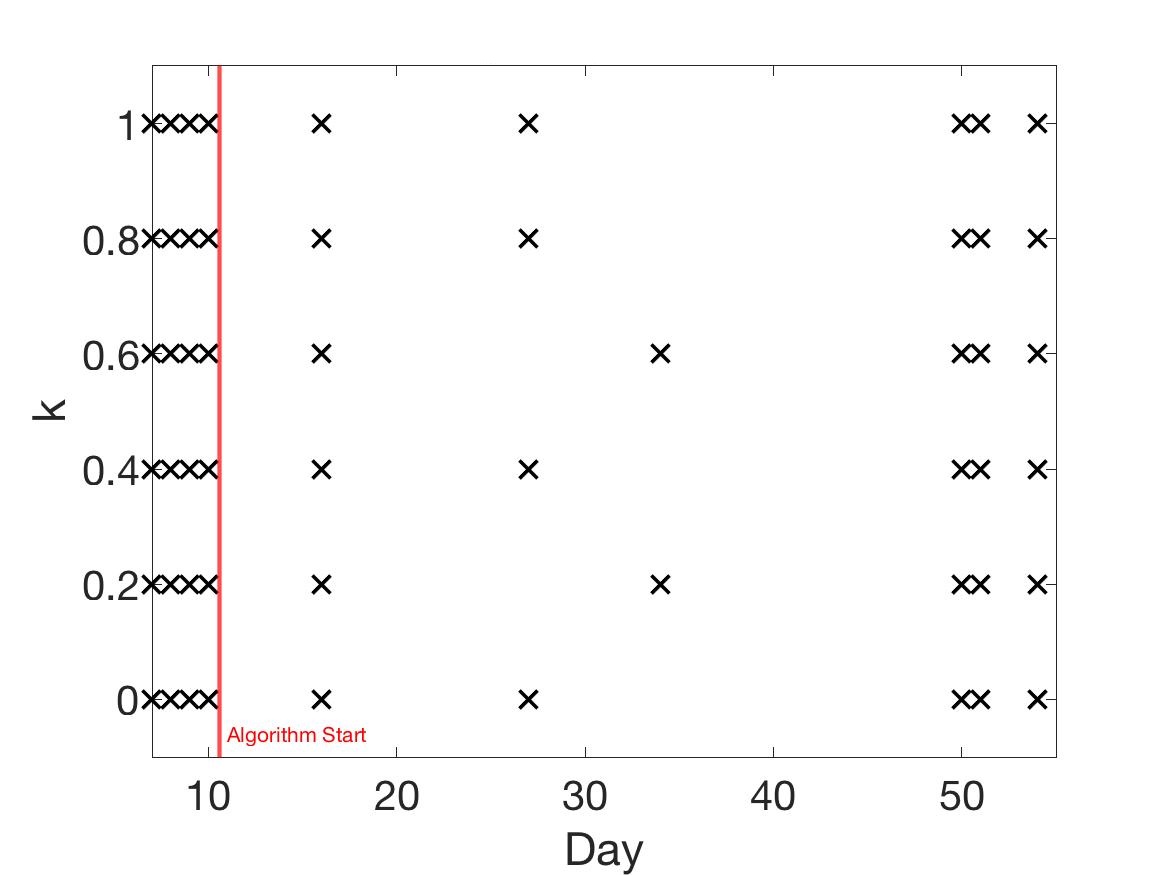}
         }
\centerline{  
         \includegraphics[width=0.49\linewidth]{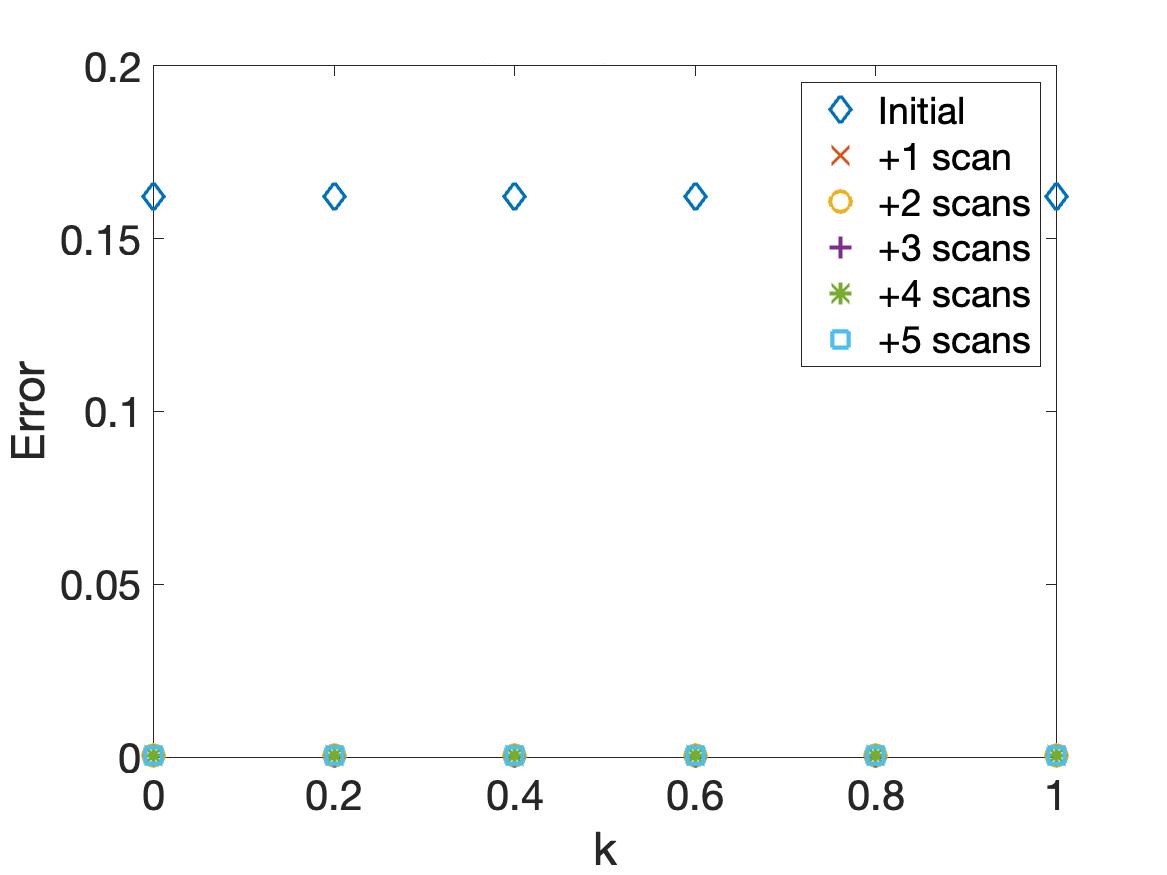}
         \includegraphics[width=0.49\linewidth]{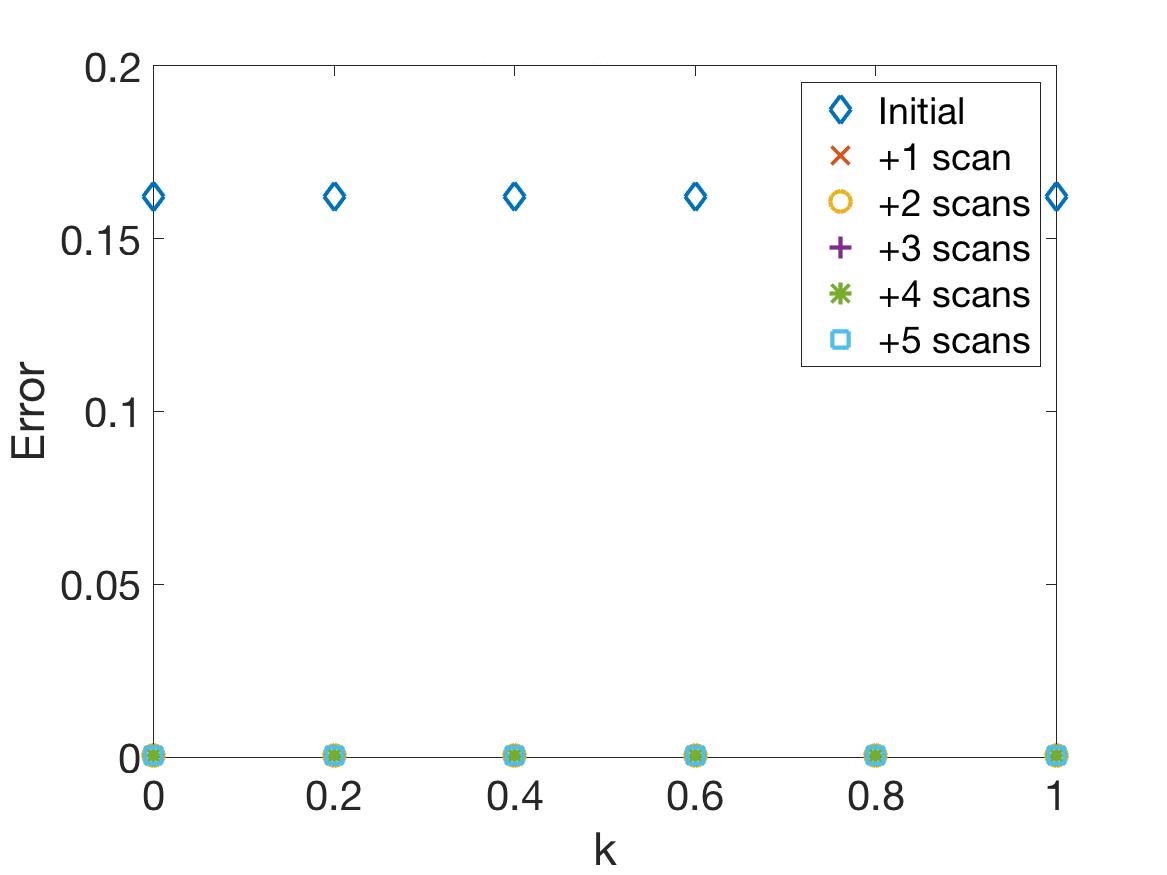}
         }
\centerline{  
         \includegraphics[width=0.49\linewidth]{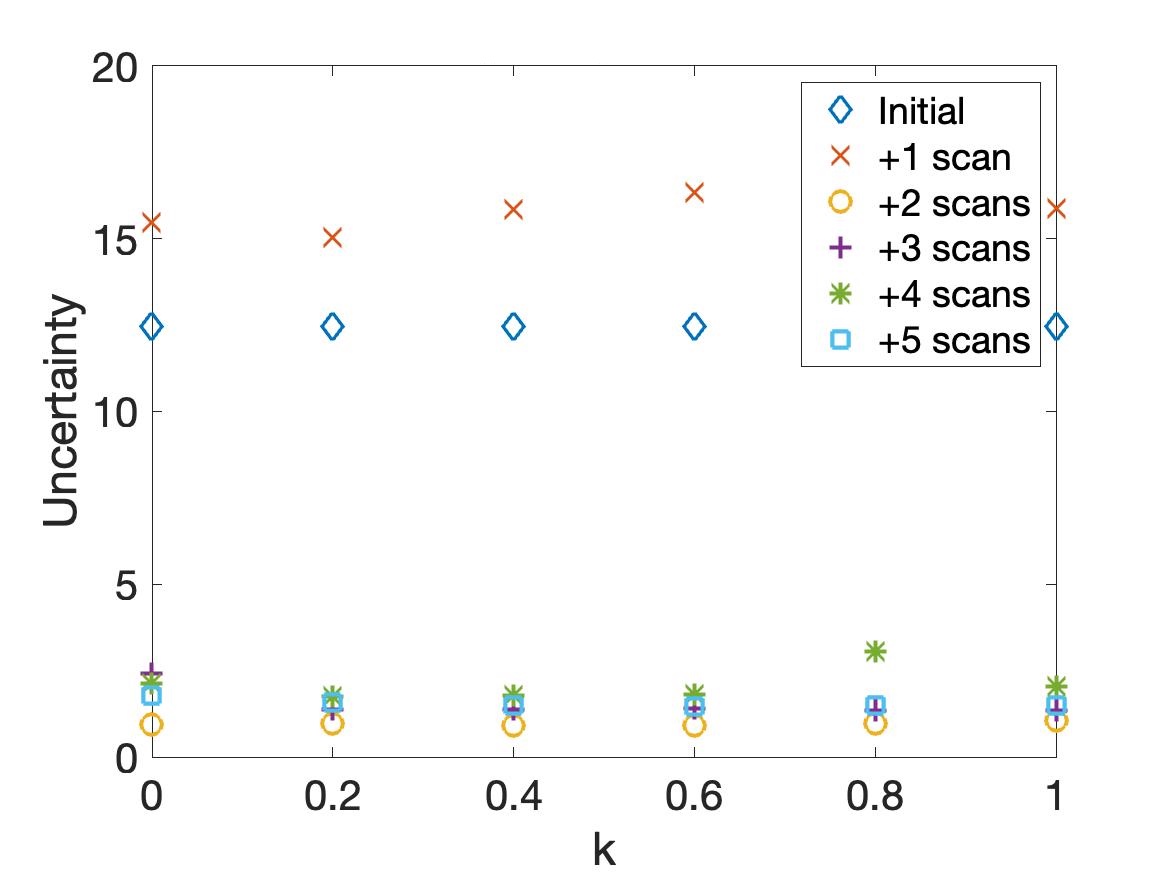}
         \includegraphics[width=0.49\linewidth]{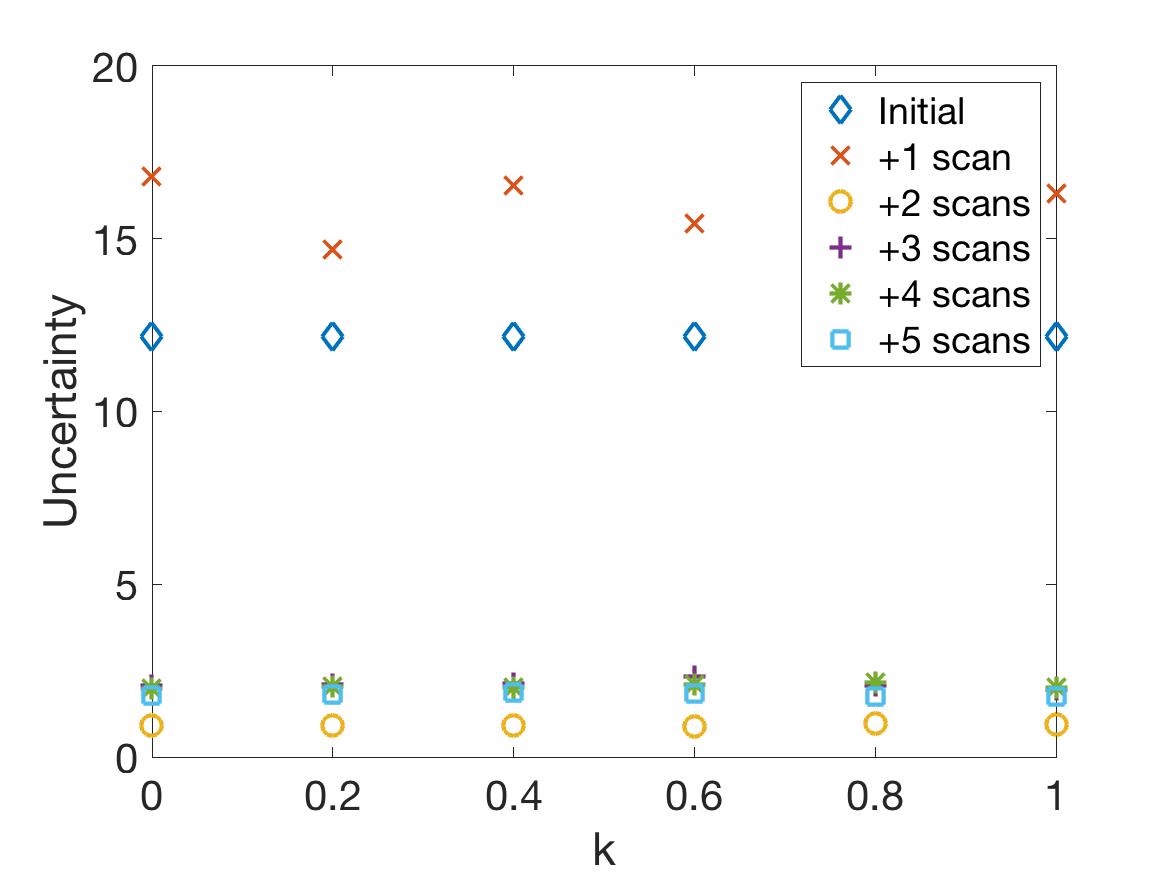}
         } 
\caption{\textit{Non-responder}. A comparison of the chosen scan choices as $k$ varies for a non-responder, using the original score function (left) and the new score function (right). The new score function chooses significantly fewer scans compared to the original score function, while achieving comparable accuracy and uncertainty levels. Moreover, the new score function gives more consistent scan schedules across $k$ values, showing less dependency upon the value of $k$. }
    \label{fig:paitient_12}
\end{figure}

Figure \ref{fig:paitient_12} displays the chosen scans for the non-responder, using the previous score function on the left, and using the new gradient-based score function on the right, in addition to the error and uncertainty. We observe that the algorithm chooses significantly fewer scans using the new score function than using the original score function. For any choice of parameter $k$ value, the first scan is chosen to be at the beginning of the treatment, and then only one additional scan is chosen within the first half of the treatment. Moreover, the plots comparing the accuracy and uncertainty between the two score functions show that using fewer scans does not detract from the accuracy of the model calibration. The error and uncertainty in the fits is comparable between the two score functions, and in both cases two scans during the treatment period is sufficient to obtain an accurate model fit. We also point out that the scan schedule is more consistent across the values of $k$ using the new score function, which shows how the scan schedule is less sensitive to the choice of parameter $k$ value, making our methodology more robust.

\begin{figure}[htbp]
\centerline{  
         \includegraphics[width=0.49\linewidth]{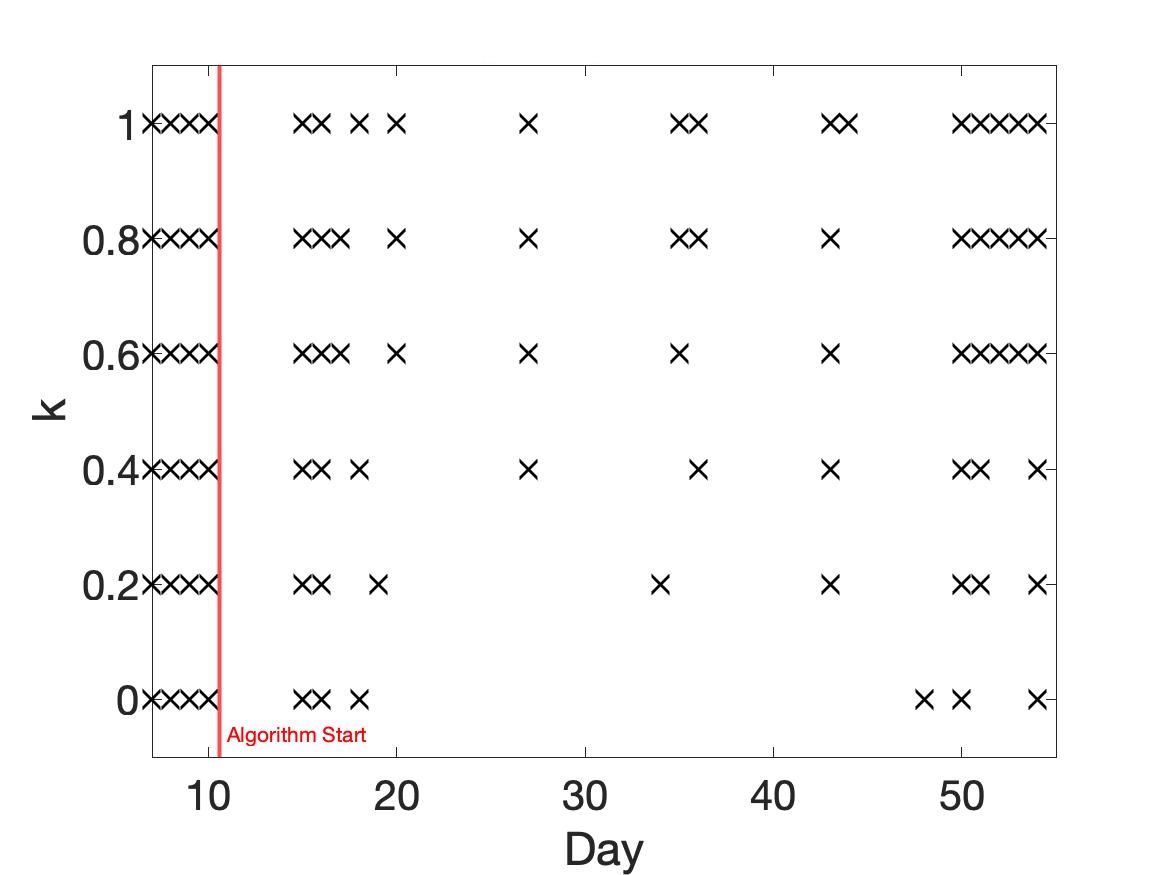}
         \includegraphics[width=0.49\linewidth]{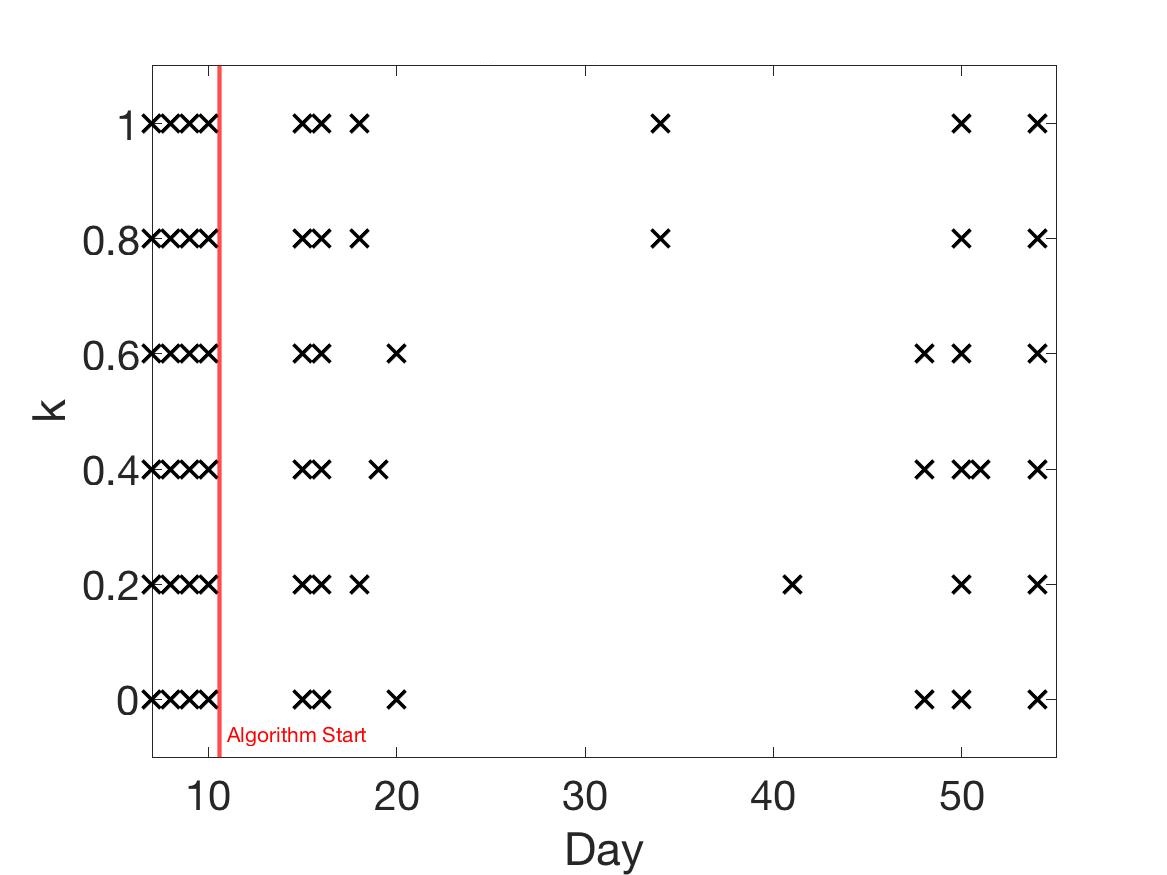}
         }
\centerline{  
         \includegraphics[width=0.49\linewidth]{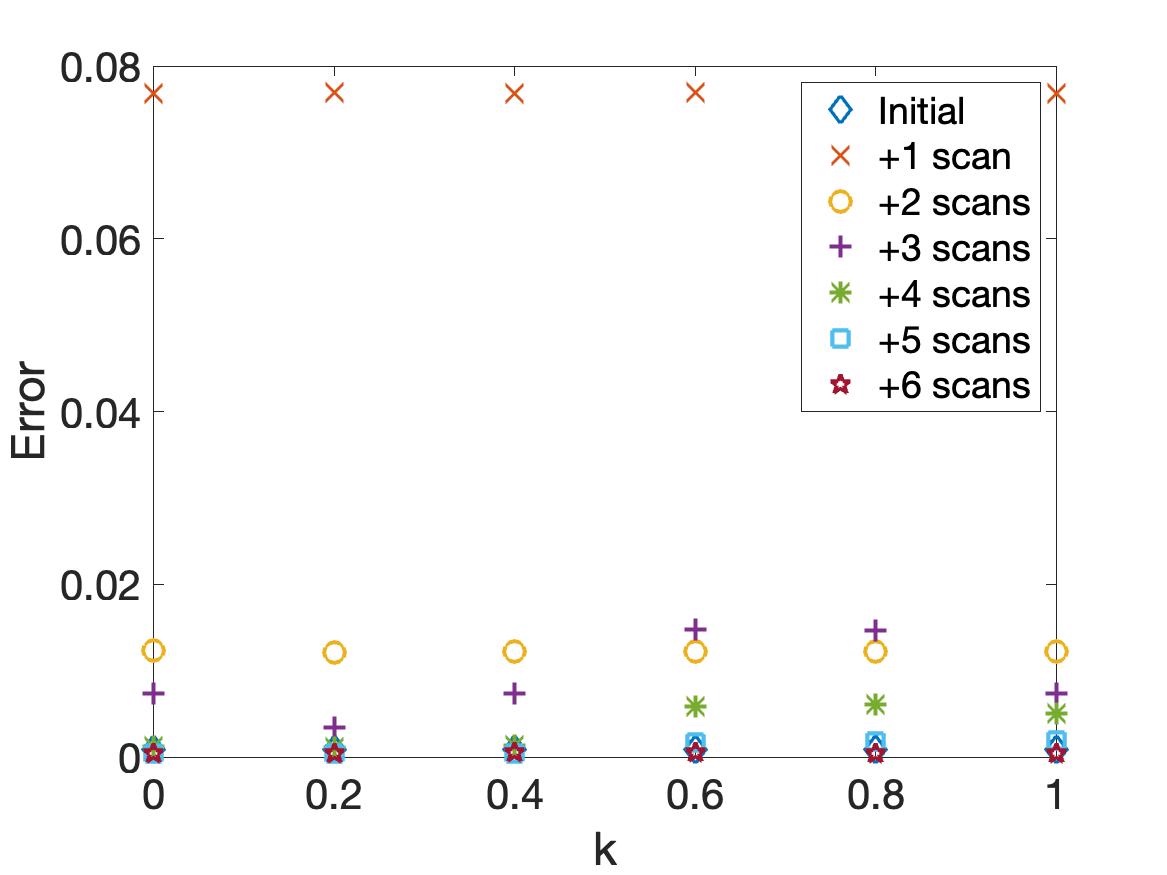}
         \includegraphics[width=0.49\linewidth]{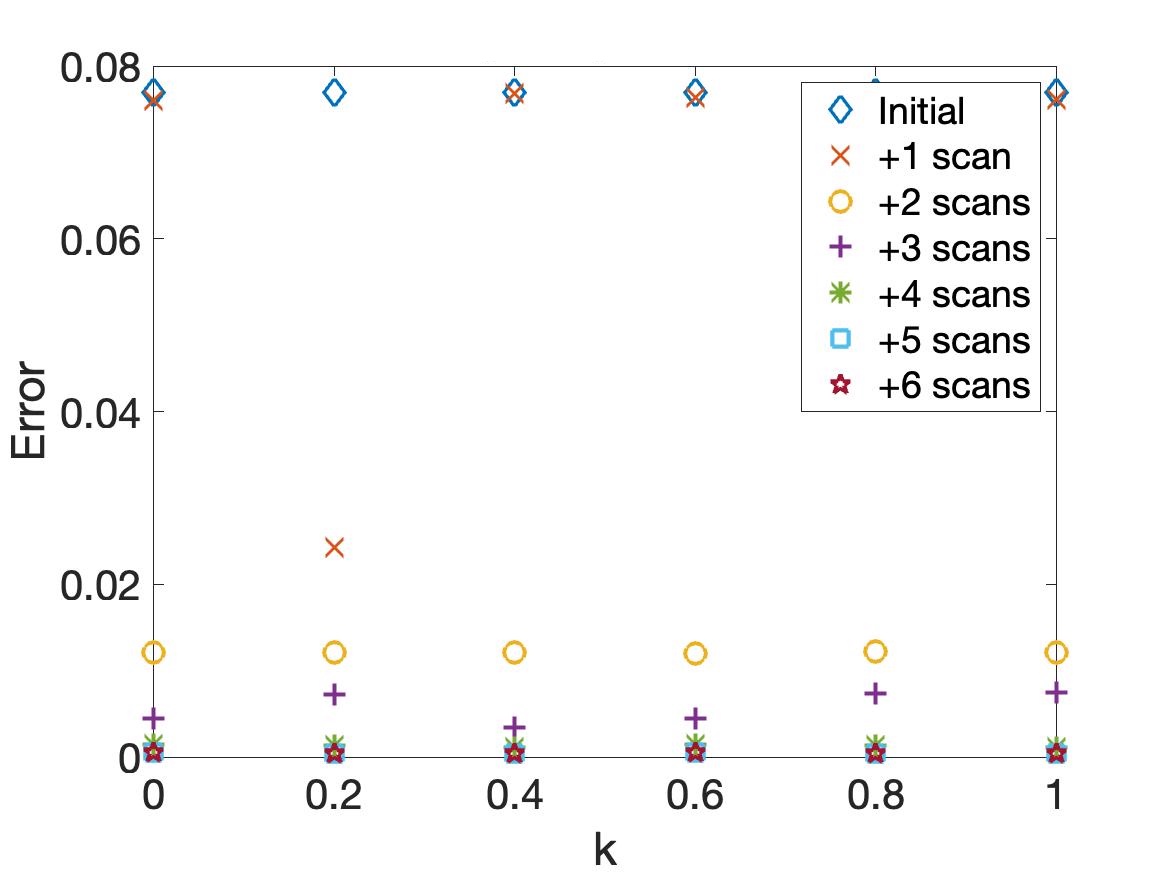}
         }         
\centerline{  
         \includegraphics[width=0.49\linewidth]{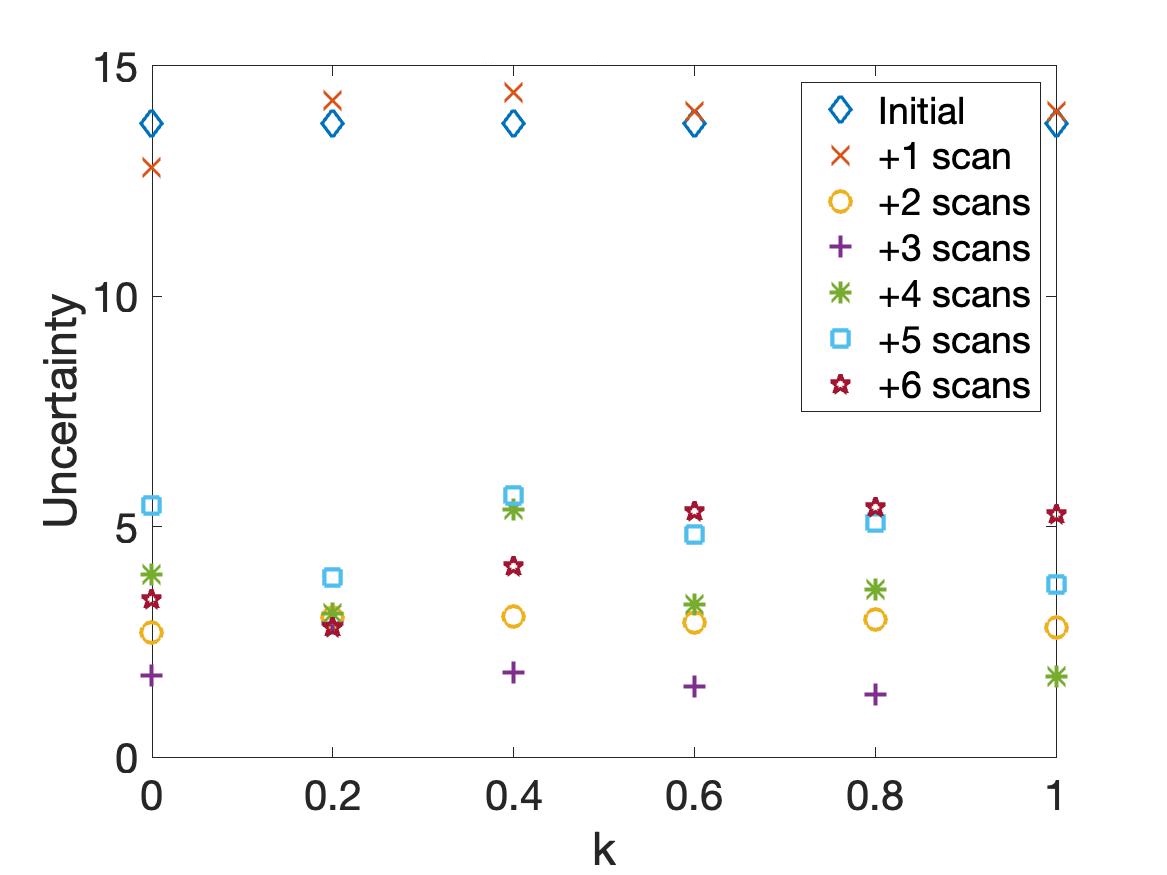}
         \includegraphics[width=0.49\linewidth]{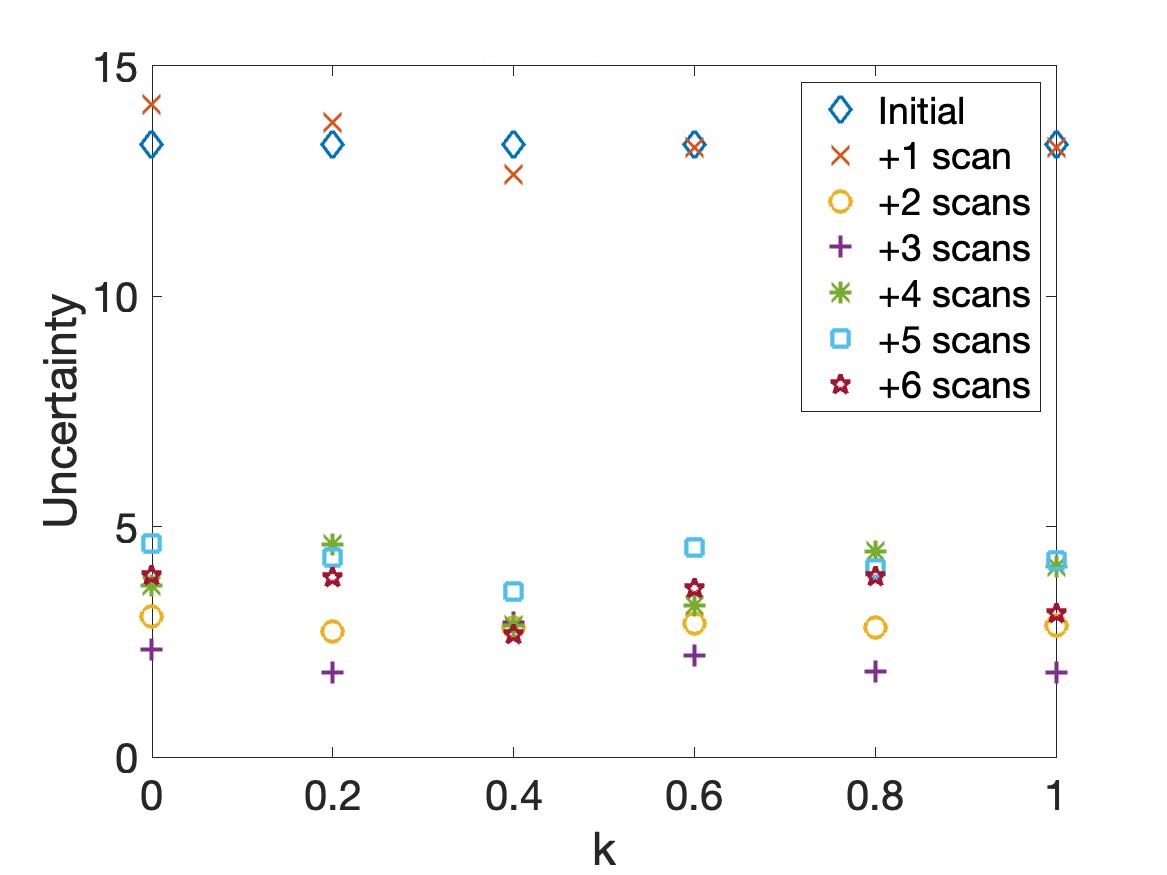}
         }         
    \caption{\textit{Medium responder}. A comparison of the chosen scan choices as $k$ varies for a medium responder, using the original score function (left), and the new score function (right).  The new score function chooses significantly fewer scans compared to the original score function, while achieving similar levels of accuracy and uncertainty. }
    \label{fig:patient_14}
\end{figure}

Figure \ref{fig:patient_14} shows the scan choices for a medium responder to radiotherapy, in which the choices on the left were made using the original score function, and the choices on the right were made using the new gradient-based score function. Similarly to the non-responder, significantly fewer scans were chosen using the new function. Figure \ref{fig:patient_14} also displays the comparison of error and uncertainty between the two score functions, for a set of scan budgets. We observe that despite the reduction in number of scans used, calibration accuracy is maintained using the gradient-based score function, as evidenced by errors and uncertainties on the same order of magnitude between the two score functions, for each scan budget.

\begin{figure}[htbp]
\centerline{  
         \includegraphics[width=0.49\linewidth]{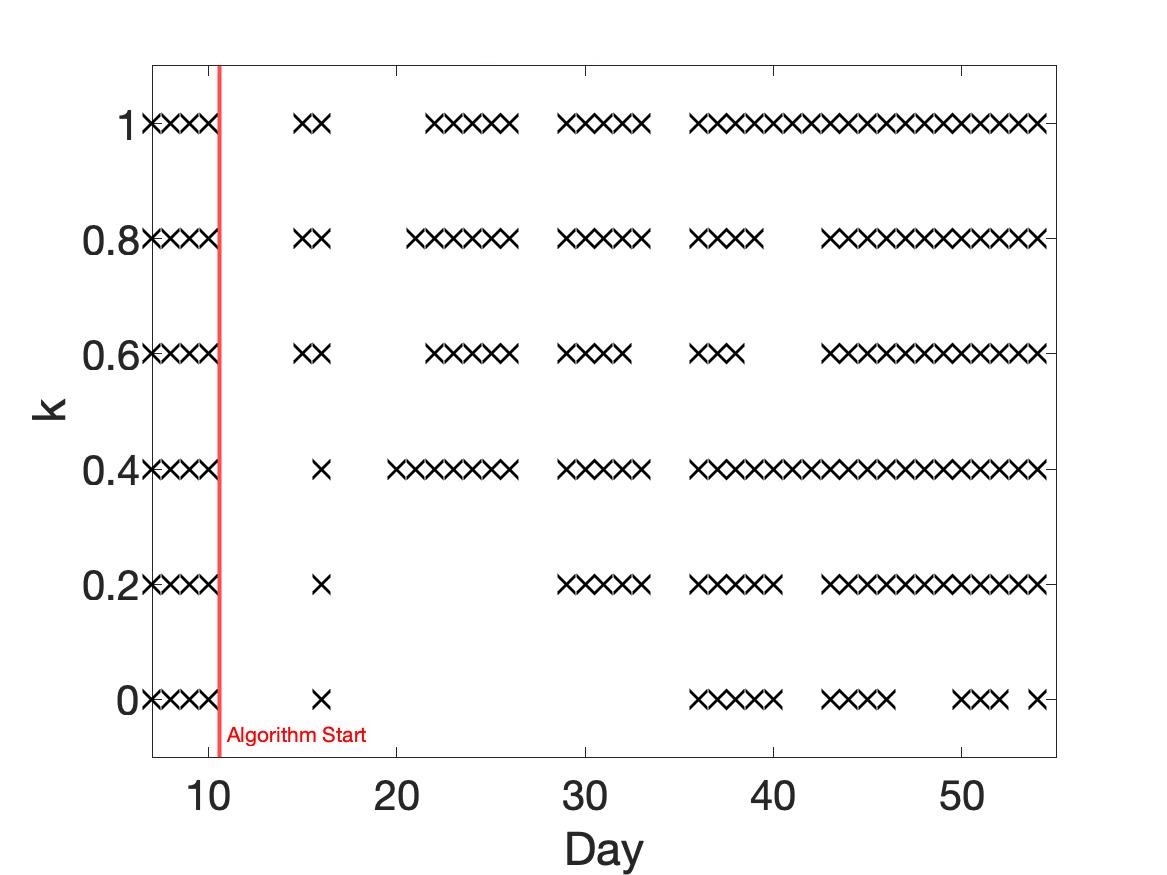}
         \includegraphics[width=0.49\linewidth]{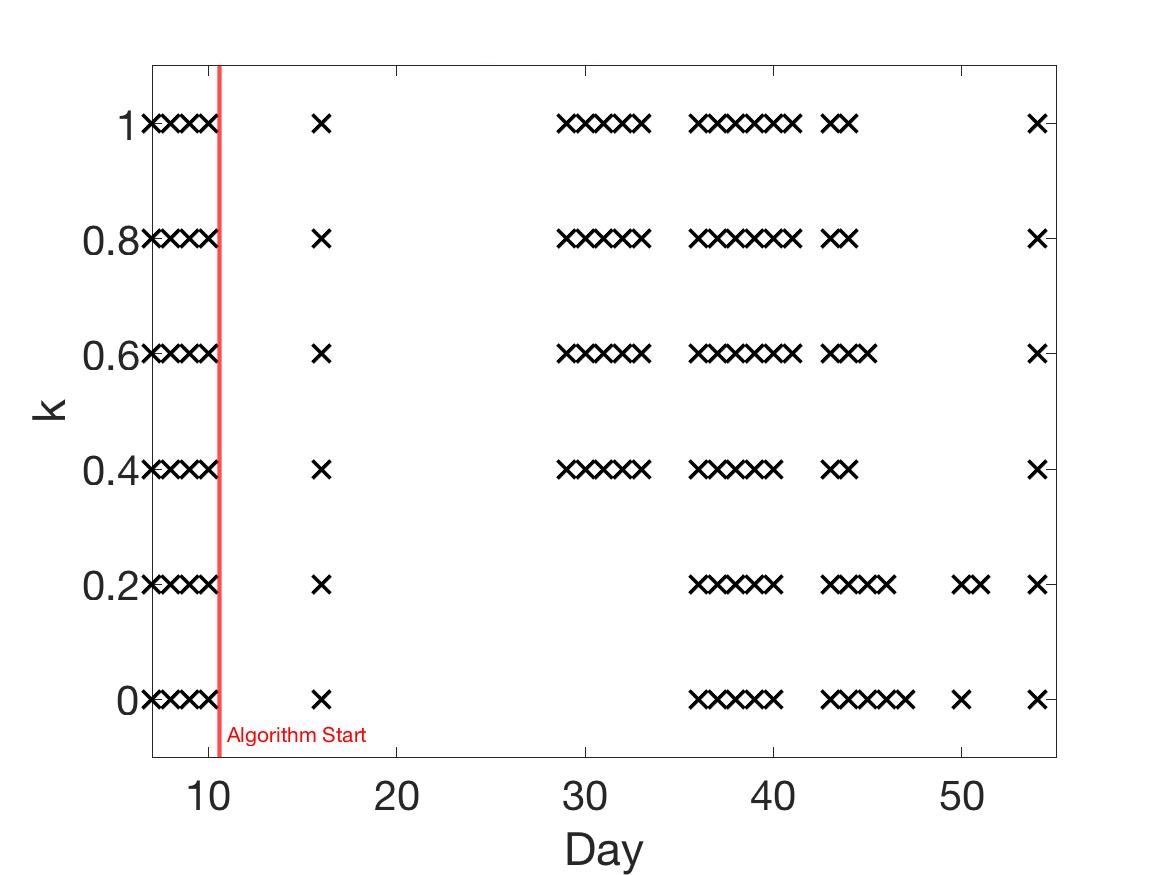}
         }
\centerline{  
         \includegraphics[width=0.49\linewidth]{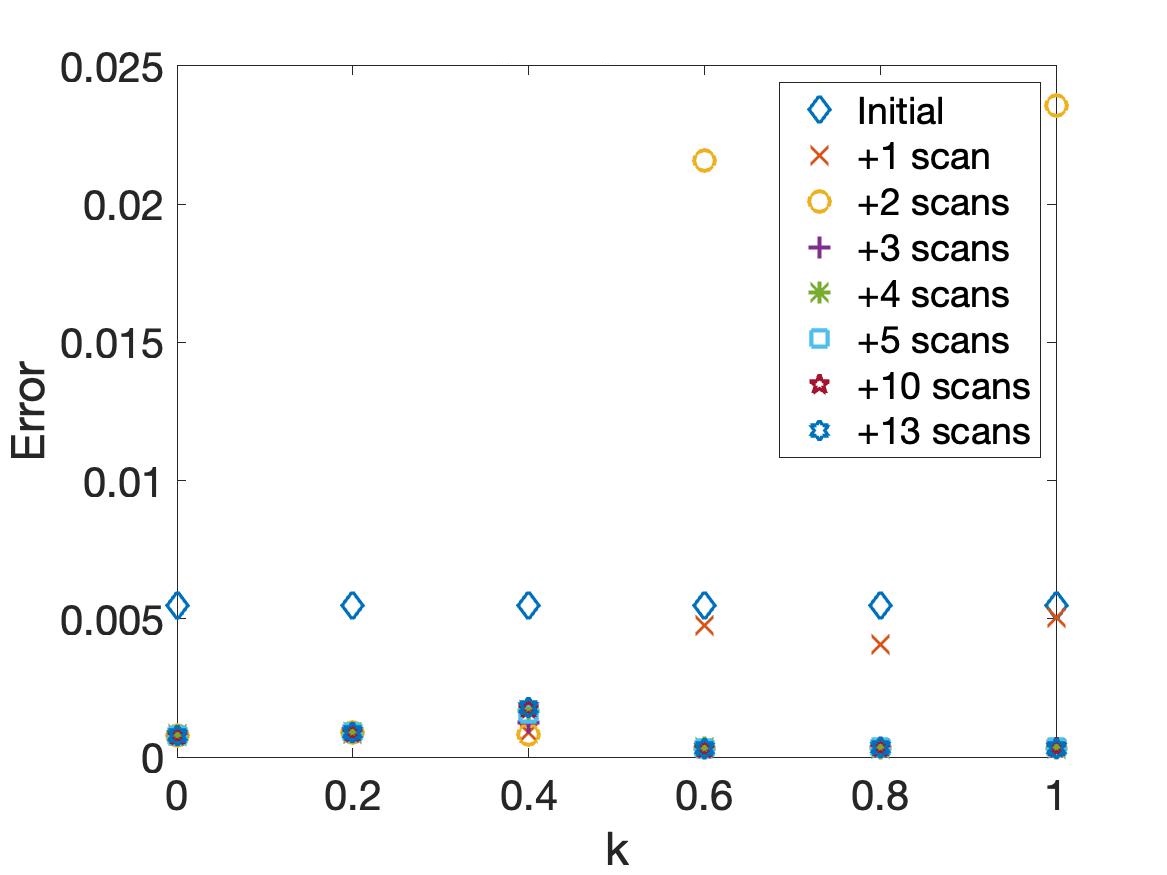}
         \includegraphics[width=0.49\linewidth]{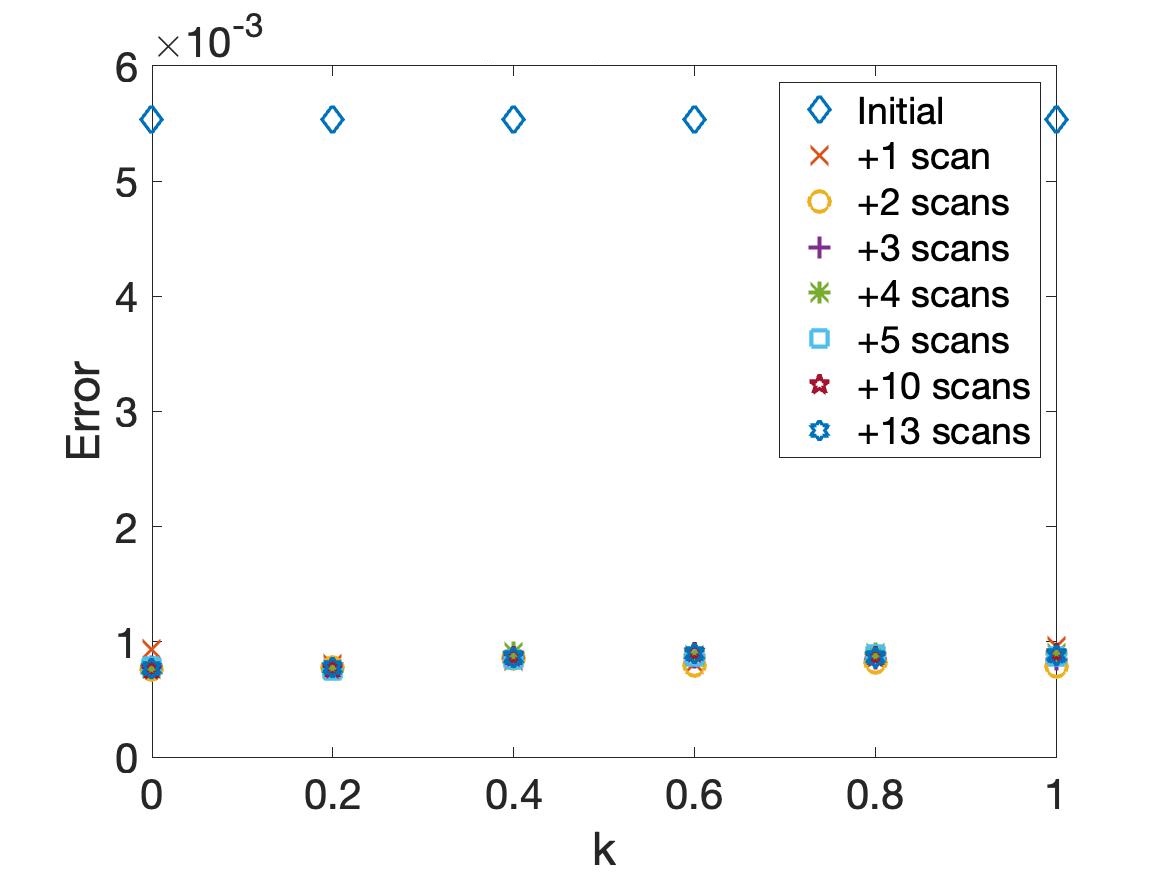}
         }
\centerline{  
         \includegraphics[width=0.49\linewidth]{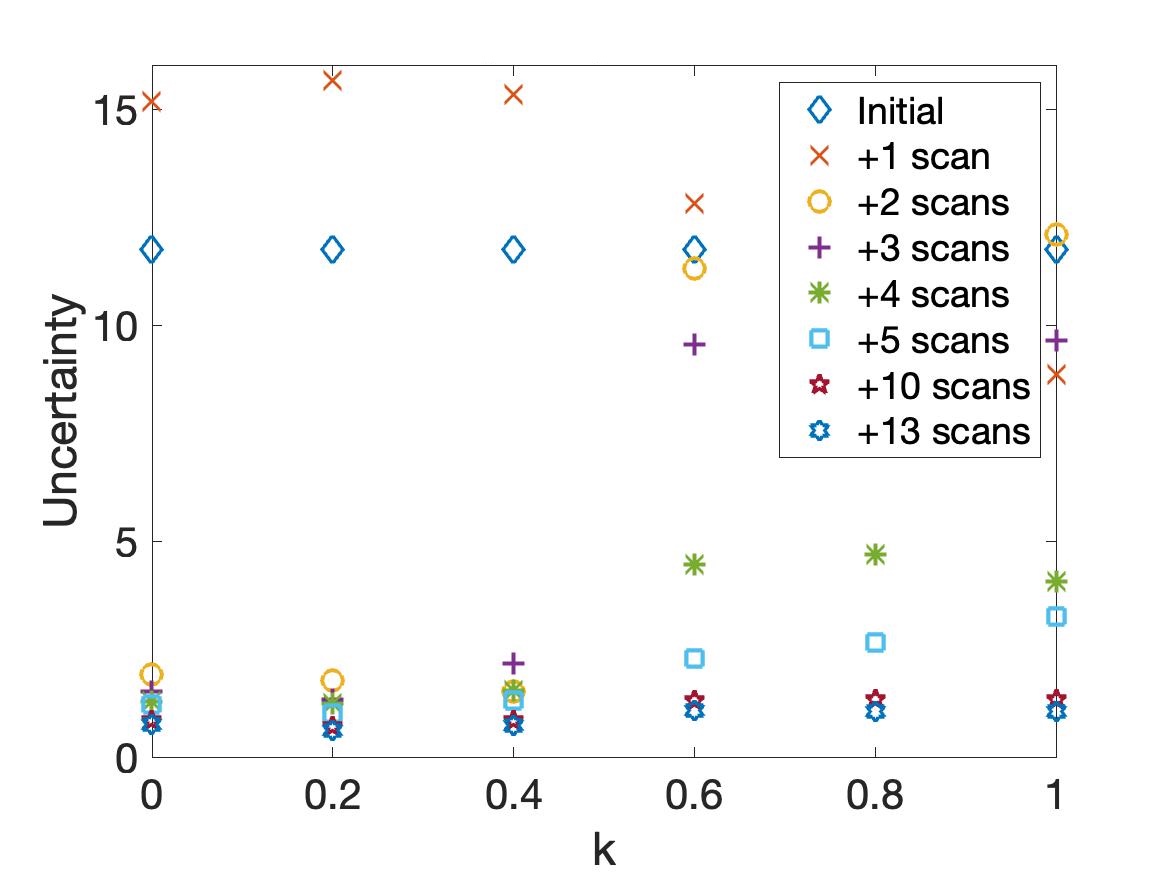}
         \includegraphics[width=0.49\linewidth]{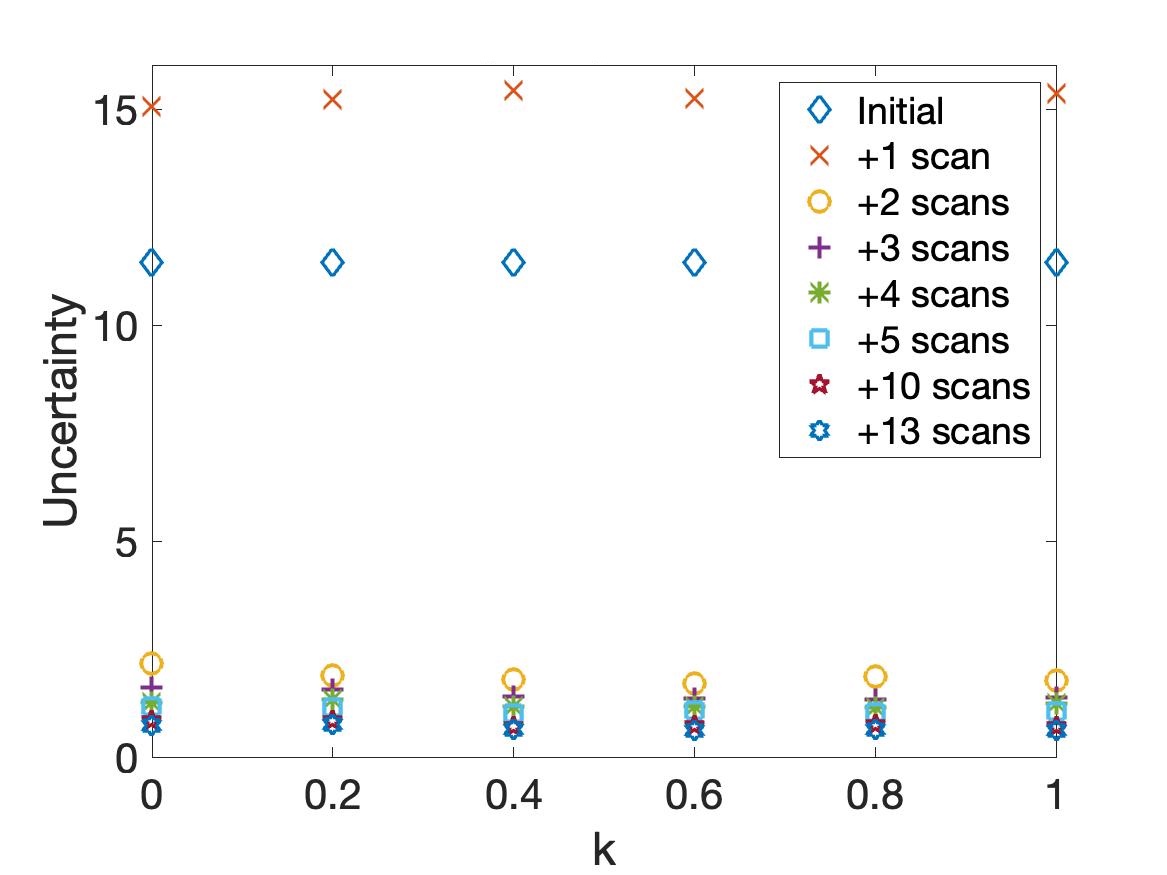}
         }
    \caption{\textit{Strong responder}. A comparison of the chosen scan choices as $k$ varies for a strong responder, using the original score function (left), and the new score function (right). The new score function fewer scans compared to the original score function, especially during the earlier portions of the treatment period. }
    \label{fig:patient_18}
\end{figure}

Figure \ref{fig:patient_18} displays the scan choices for a strong treatment responder. Here, the trend observed for the non-responder and medium responder continues, i.e.~the number of chosen scans is smaller using the gradient-based score function than using the previous score function.  The error and uncertainty comparisons between the two score functions are shown in Figure \ref{fig:patient_18} as well; we see that using the new score function, just one scan after treatment begins on day 15 is sufficient to accurately predict the treatment response. Just as in the previous cases, the error and uncertainty for each scan budget is comparable between the two score functions.

An important difference between the two score functions is that the error and uncertainty are much more stable across the $k$ values when we use the new gradient-based score function than when we use the original score function. The stability difference between the two score functions is most pronounced for the strong responder, but the error and uncertainty are also reasonably stable using the new score function in the non-responder and medium responder cases. Thus, it is reasonable to eliminate the parameter $k$ and set a fixed value for $k$ in our gradient-based score function. We choose $k=1$ because this parameter choice favors scans that are collected earlier in time, which aligns with our goal to accurately calibrate the model as soon as possible after treatment begins. The ability to eliminate the parameter $k$ is an important benefit of the gradient-based score function, since the optimal $k$ value varies from patient to patient when using the previous score function, making it much less broadly applicable than the gradient-based score function.

\subsection{Assessment of Model Accuracy and Uncertainty}

Having determined that the penalization parameter $k$ can be set to $k=1$ within the proposed score function of Section \ref{sec:newscorefxn}, we now analyze the results of the calibration procedure for each of our three sample patients, using the following score function, updated by fixing $k=1$,

\begin{equation}
S(i,r) = R(i,r)-\bigg \lvert \frac{\tilde d_r-\tilde d_{r-1}}{\tilde d_{r-1}(t_r-t_{r-1})}\bigg \rvert \left(\frac{\sum\limits_{j=r+1}^{i-1} R(j,r)}{\sum\limits_{l=r+1}^{n_T}R(l,r)}\right). 
\label{eq:NewScore_nok} 
\end{equation}

For each case, we consider how the fits of the model trajectory and the width of the associated credible intervals evolve as more data points are added to the collection, and use the uncertainty and error metrics discussed in Section \ref{sec:assessment} to determine how many total scans are needed to meet our accuracy and uncertainty goals. In order to fully assess how the algorithm works, we allow it to run without any budget constraints, so that it will continue to choose scans until it runs out of options on day 54.  However, we note that in many cases, inclusion of all these scans is not necessary, and we consider how to determine whether the algorithm may be terminated early.

\begin{figure}[htbp]
\centerline{ 
         \includegraphics[width=0.29\linewidth]{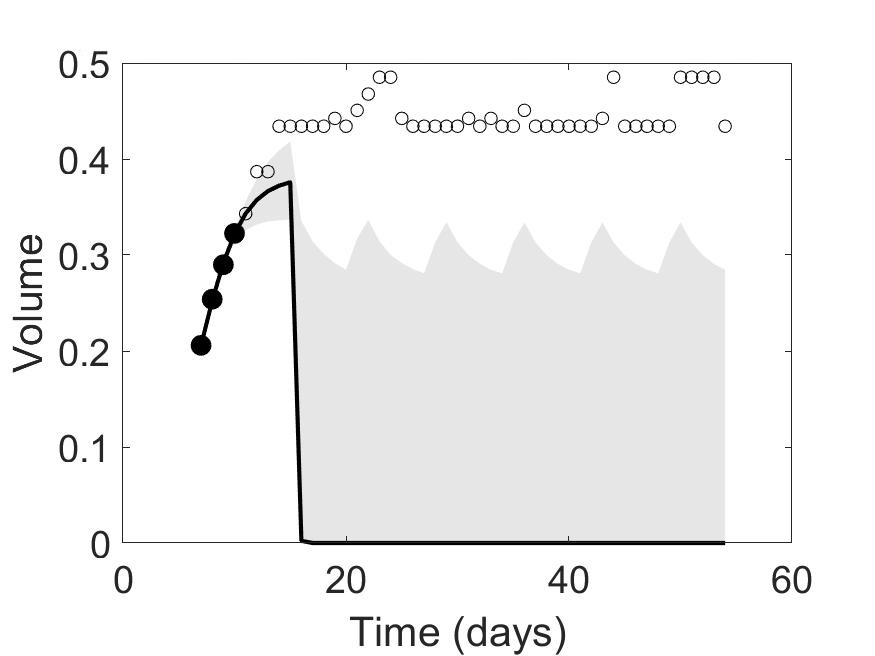}
         \includegraphics[width=0.29\linewidth]{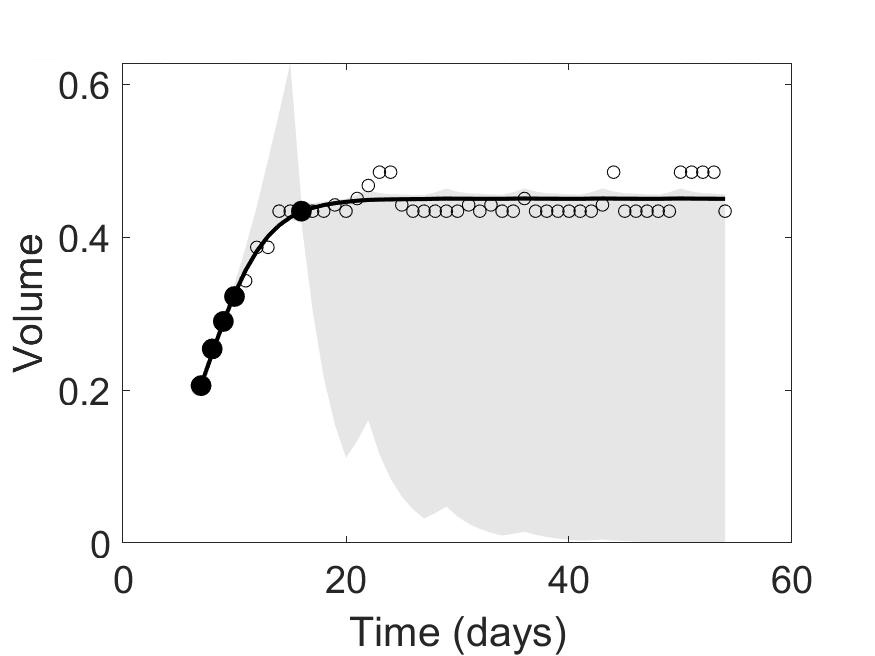}
         \includegraphics[width=0.29\linewidth]{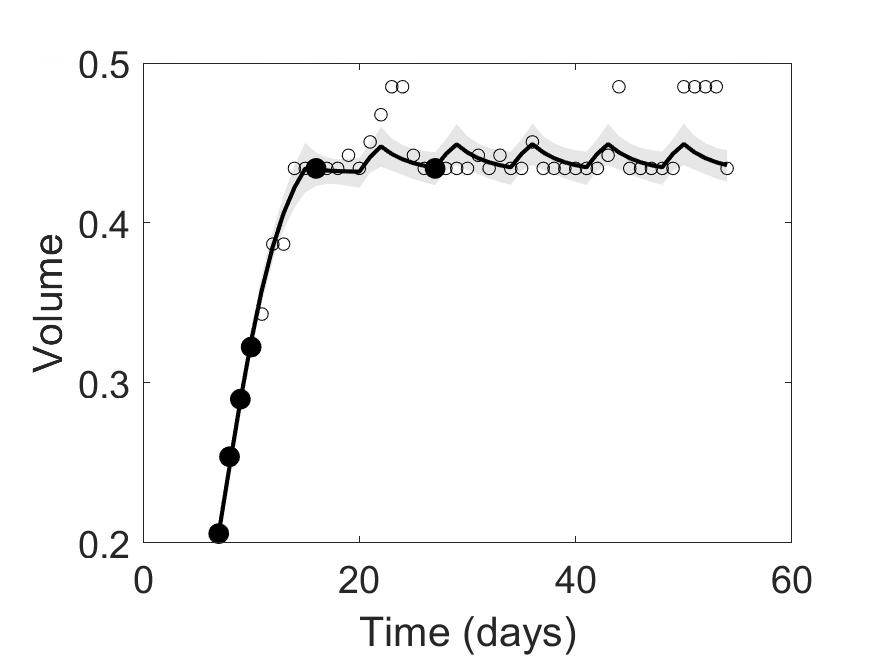}
}
\centerline{ 
        \includegraphics[width=0.29\linewidth]{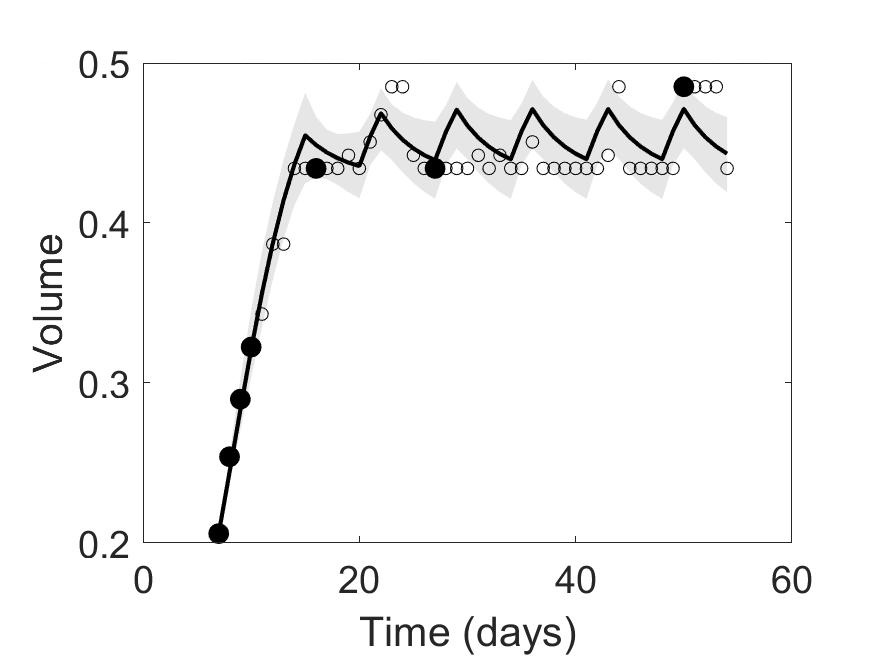}
         \includegraphics[width=0.29\linewidth]{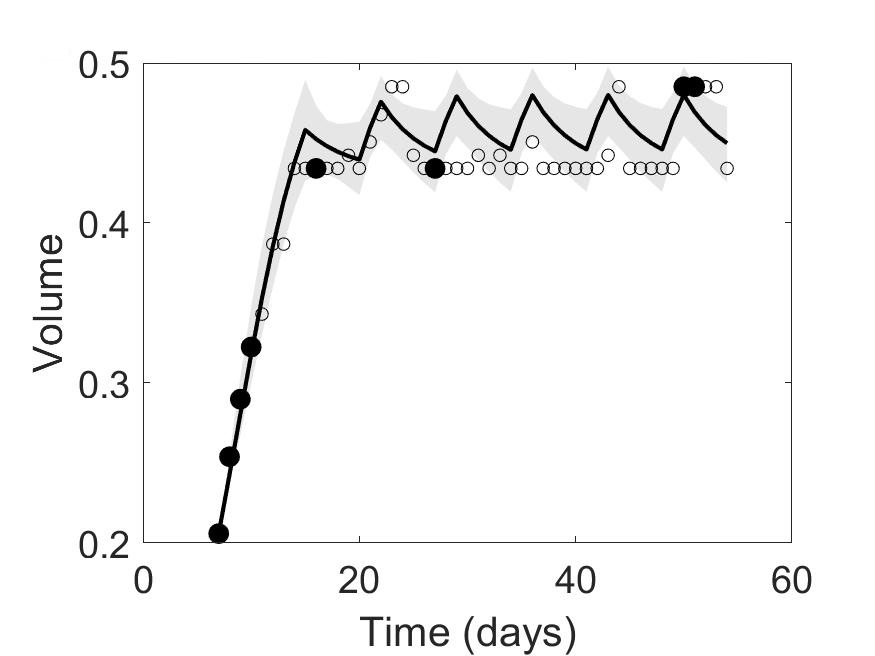}
         \includegraphics[width=0.29\linewidth]{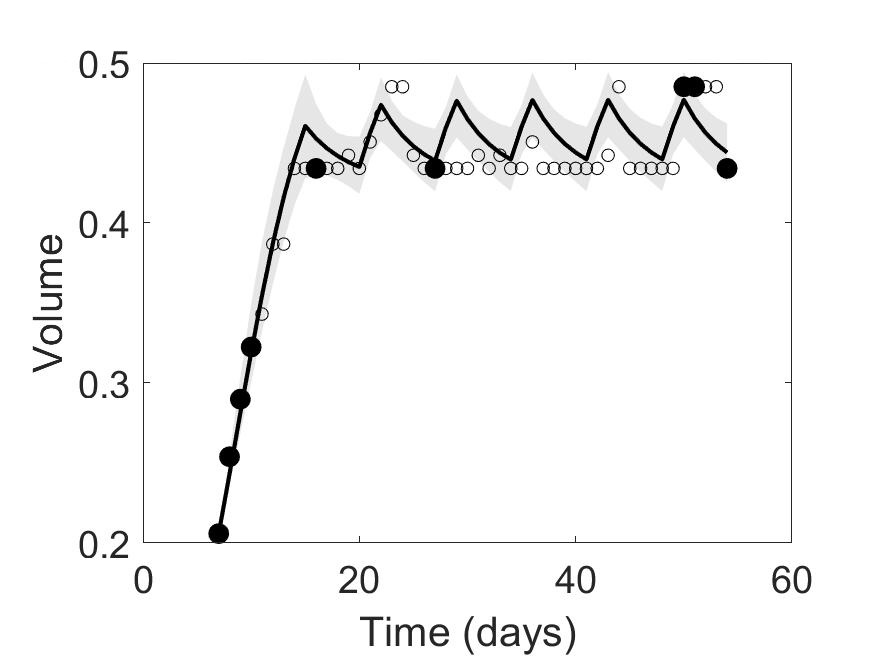}
}
    \caption{\textit{Non-responder} Credible interval evolution over scan progression. The first plot shows the initial calibration using four pre-treatment data points. The subsequent plots show the progression as the following five scans are added. We observe that the credible interval narrows down significantly after adding two scans (which agrees with the uncertainty shown in Figure \ref{fig:paitient_12}.) } 
    \label{fig:credintprog12}
\end{figure}

We begin with our non-responder spheroid, with calibration results shown in Figure \ref{fig:credintprog12}. After the initial seeding of the algorithm with data from days 7-10---and with no constraints given on budget---the algorithm selects scans on days 16, 27, 50, 51, and 54. At each step, the posterior distributions for the parameter set $[A, B, \beta]$ are fed through the model to generate a 95\% credible interval about the output trajectory; this is shown as gray shading in the model fit figures. Our initial fit and the fit resulting from inclusion of day 16 both display a large amount of model uncertainty, resulting from the fact that little to no data from the treatment period has yet been supplied.  Once we add a scan on day 27, the credible interval tightens sharply; enough information has now been supplied for the algorithm to guess that this spheroid will not undergo drastic changes in volume over the treatment period. From this point forward, both the model trajectory and credible interval areas are relatively stable.  It becomes apparent that not all of the chosen scans are necessary, and we could choose to terminate the algorithm early.

\begin{figure}[htbp]
\centerline{ 
         \includegraphics[width=0.29\linewidth]{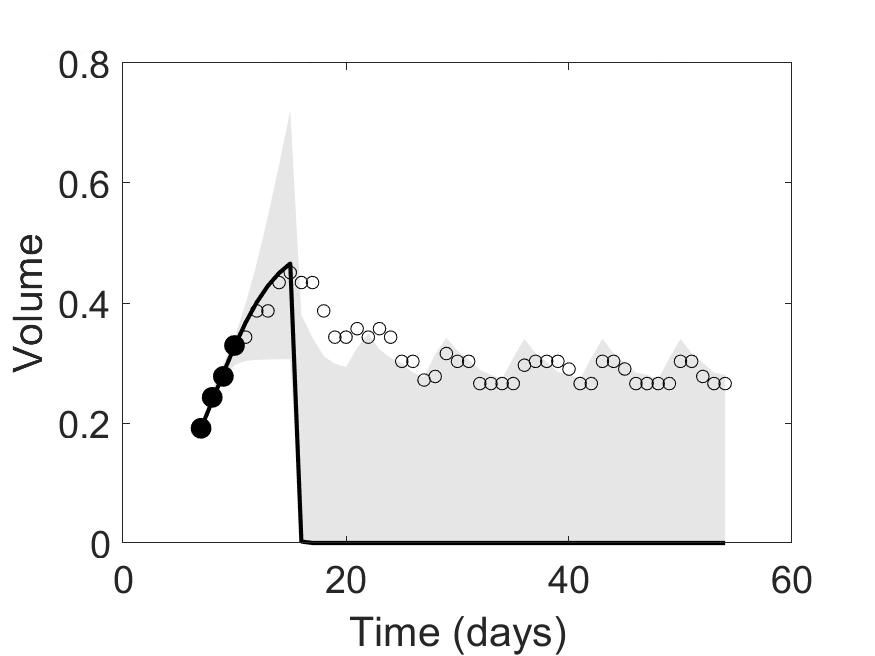}
         \includegraphics[width=0.29\linewidth]{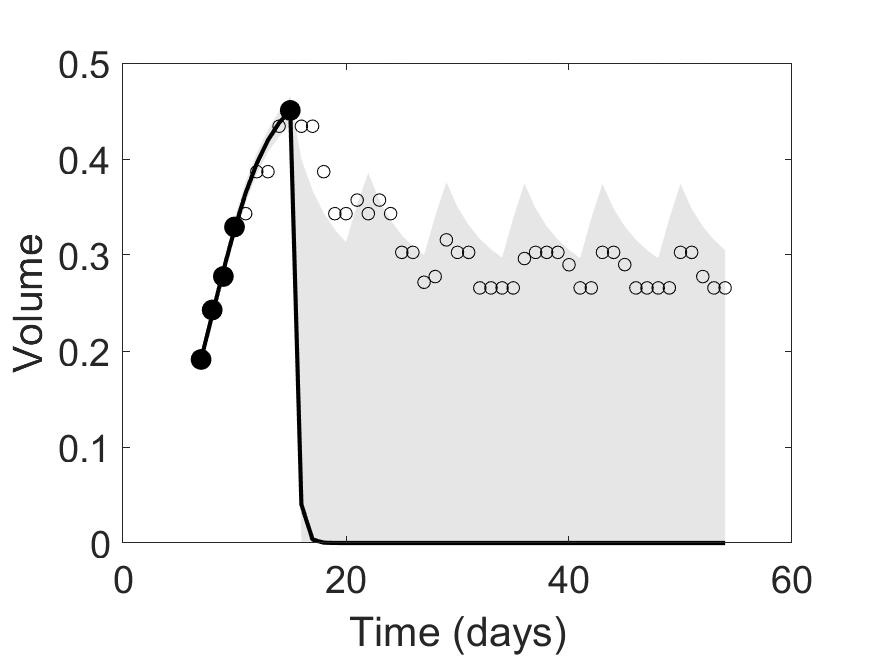}
         \includegraphics[width=0.29\linewidth]{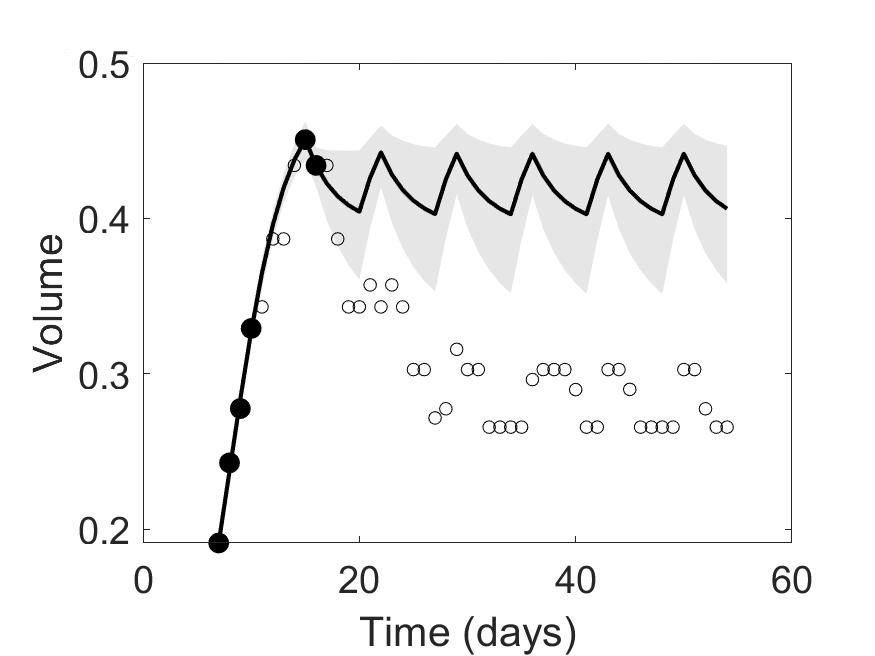}
}
\centerline{ 
         \includegraphics[width=0.29\linewidth]{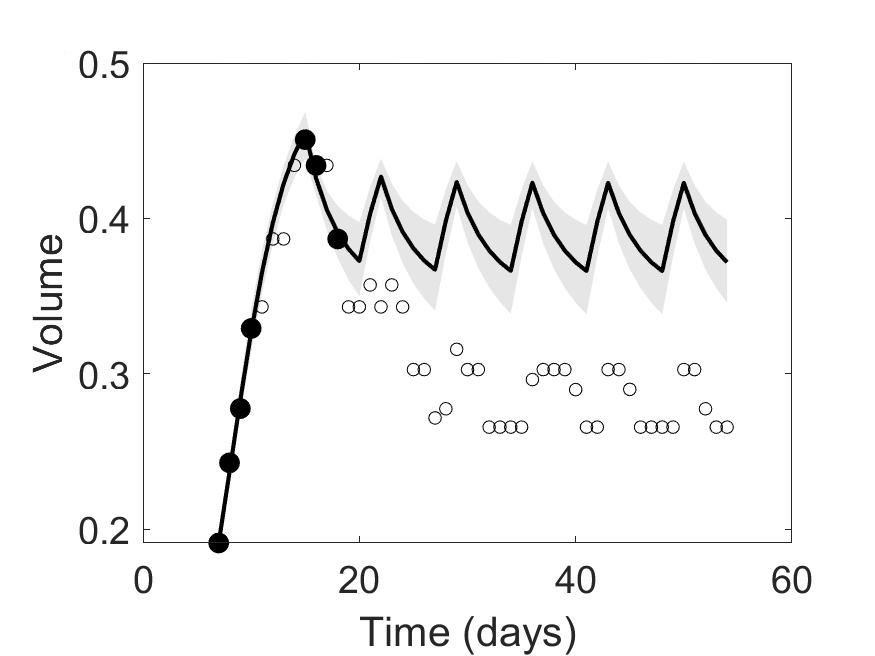}
         \includegraphics[width=0.29\linewidth]{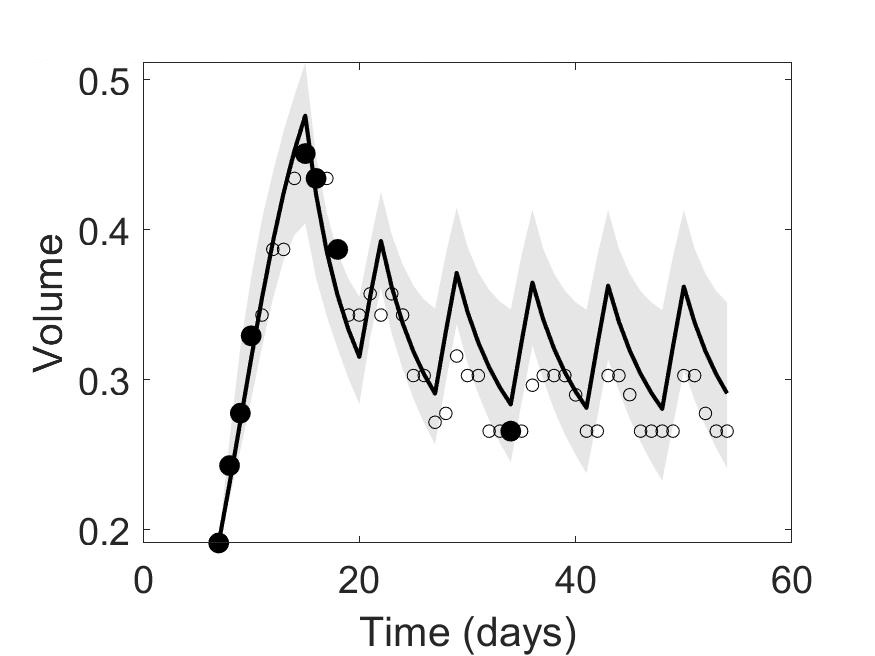}
         \includegraphics[width=0.29\linewidth]{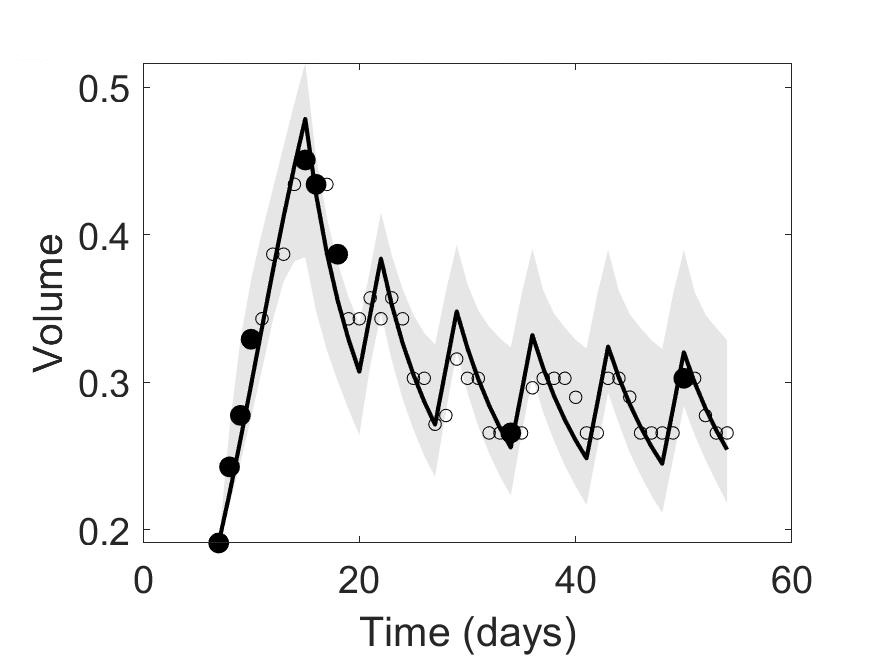}
}
    \caption{\textit{Medium responder} Credible interval evolution over scan progression. The first plot shows the initial calibration using four pre-treatment data points. The subsequent plots show the progression as the remaining scans are added. We observe that while the credible interval tightens after just two additional scans, the trajectory has not stabilized until the inclusion of four additional scans. 
    }
    \label{fig:credintprog14}
\end{figure}

The model trajectory and credible interval progression for the medium responder spheroid is shown in Figure \ref{fig:credintprog14}.  Here, the algorithm selects scans from days 15, 16, 18, 34, 50, and 54. As before, the tightening of the credible interval does not occur until two scans from within the treatment period have been supplied.  Here, we observe how to balance two goals: while it may seem that the uncertainty has been fully reduced by day 18, our model trajectory is still widely variable from scan-to-scan until inclusion of the day 34 scan.  Particularly for spheroids that exhibit a steep growth or reduction over a short period of time, it should be determined that the model trajectory has stabilized prior to terminating the algorithm, even if the uncertainty appears to have been reduced to an acceptable amount.   

\begin{figure}[htbp]
\centerline{ 
         \includegraphics[width=0.29\linewidth]{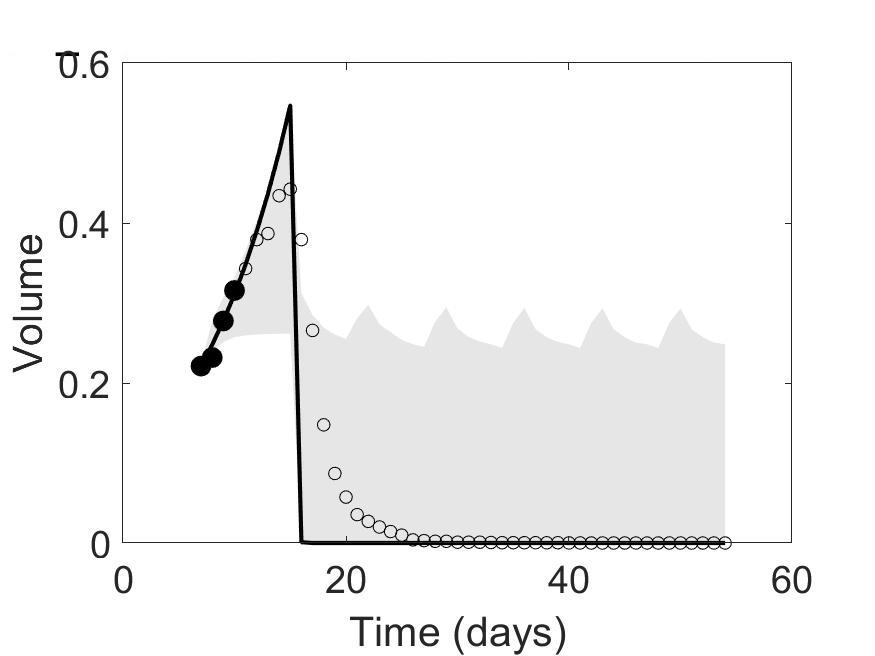}
         \includegraphics[width=0.29\linewidth]{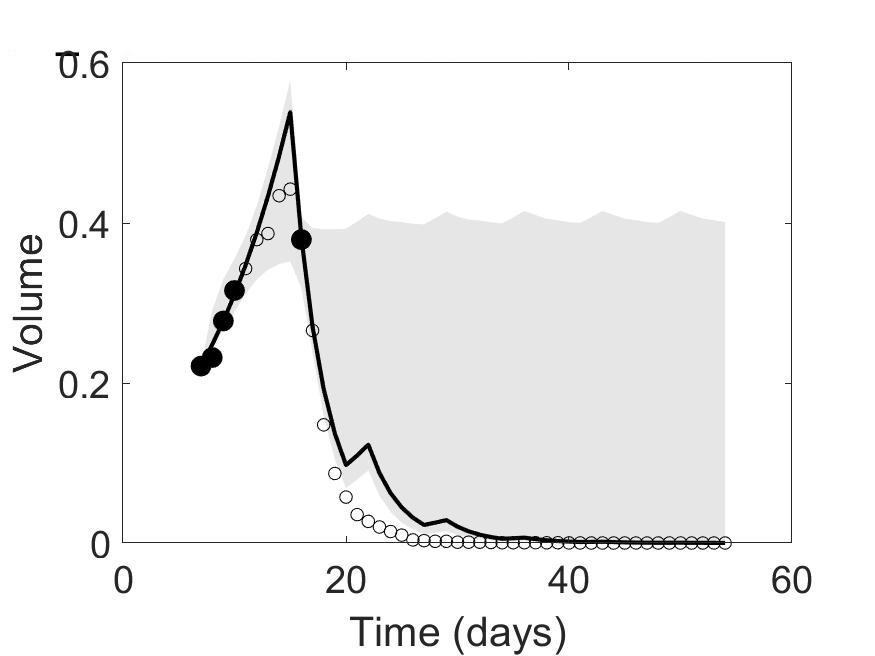}
         \includegraphics[width=0.29\linewidth]{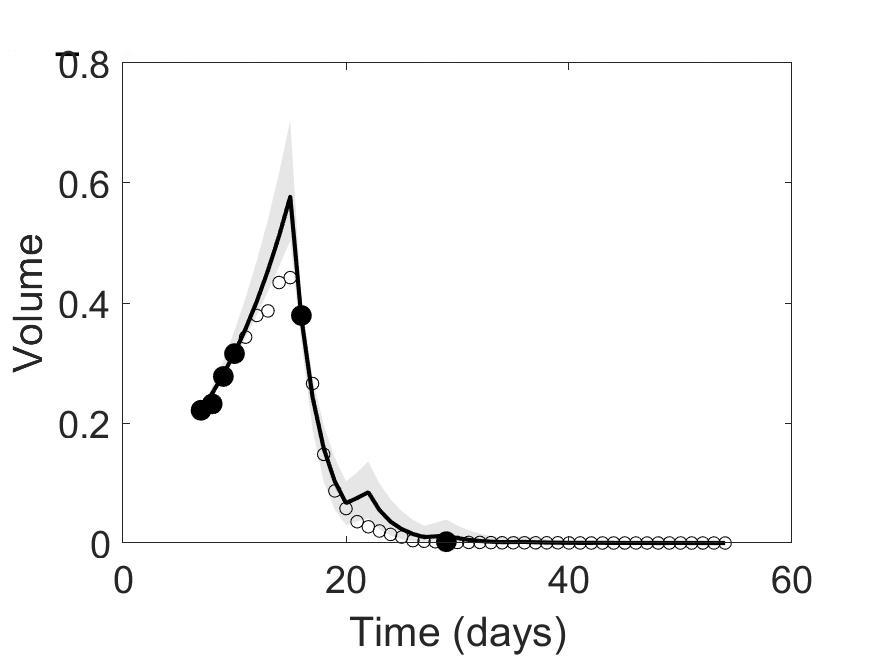}
}
\centerline{ 
         \includegraphics[width=0.29\linewidth]{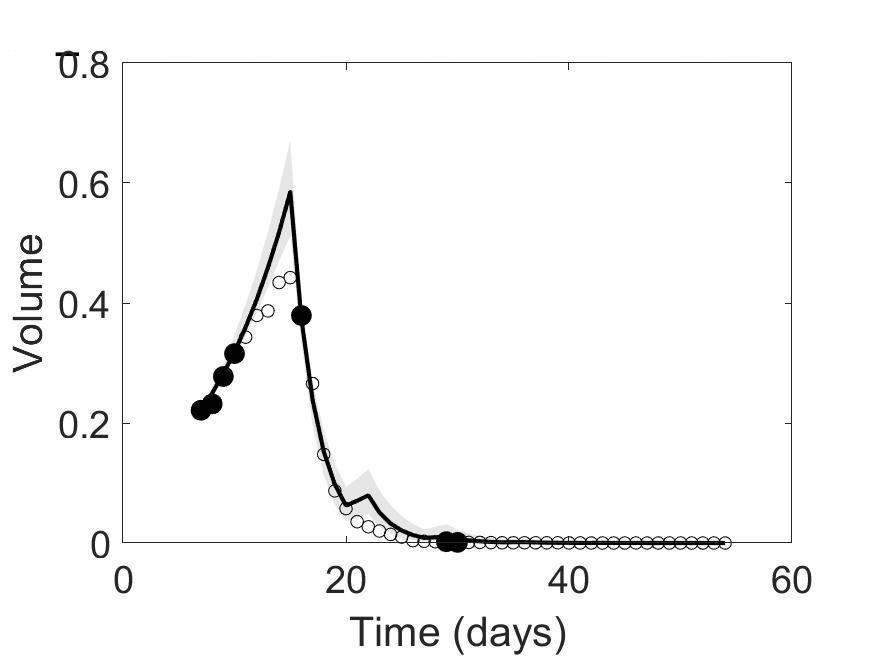}
         \includegraphics[width=0.29\linewidth]{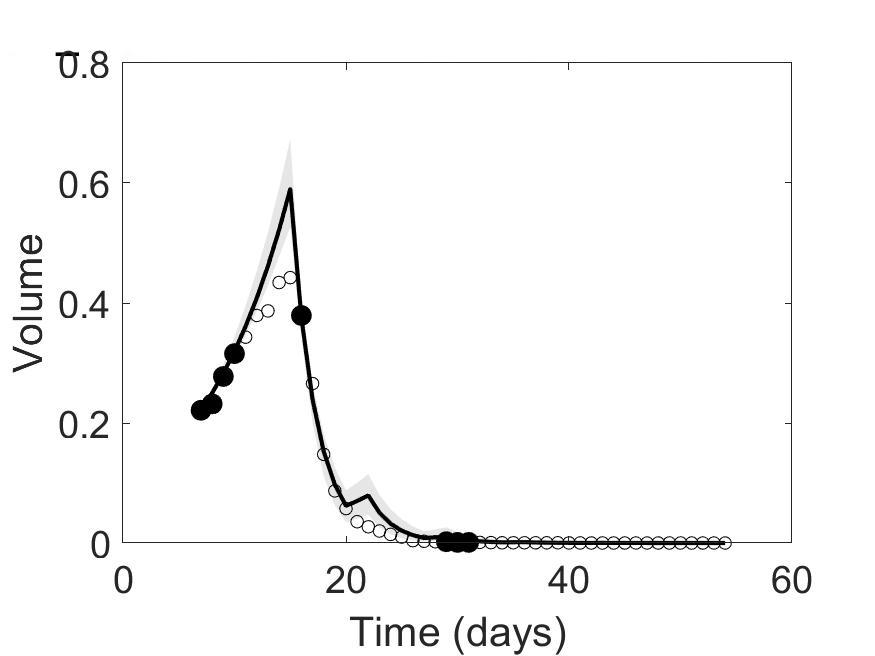}
         \includegraphics[width=0.29\linewidth]{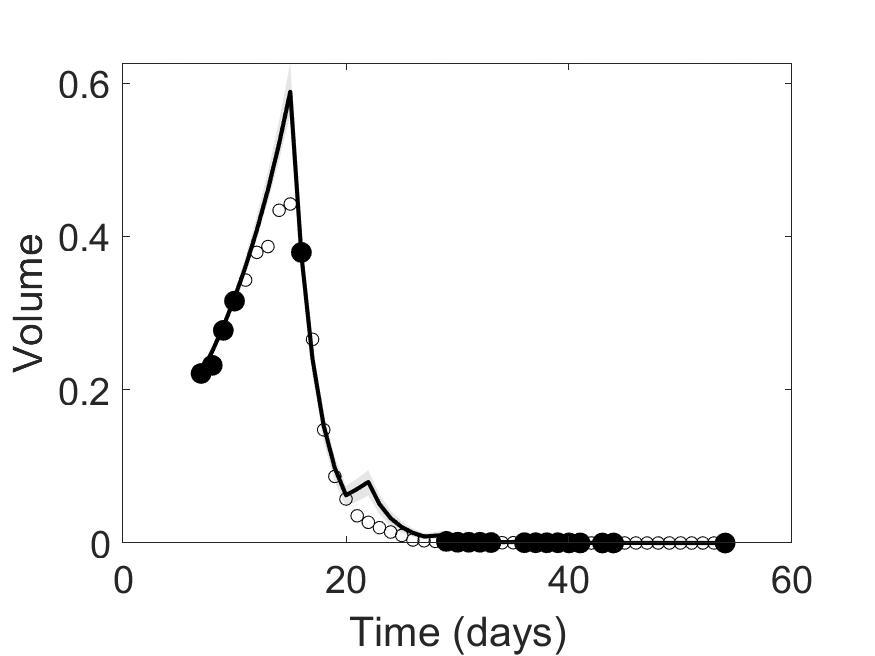}
}         
    \caption{\textit{Strong responder} Credible interval evolution over scan progression. The first plot shows the initial calibration using four pre-treatment data points. The subsequent plots show the progression when scans 1, 2, 3, 4, and 16 are added. The predicted trajectory and credible interval becomes accurate after two scans. All scans after the first two additions are unnecessary as the tumor volume is already reduced significantly. We note that a threshold could be appended to the algorithm to stop the scanning protocol once the tumor volume is below such a threshold. }
    \label{fig:credintprog18}
\end{figure}

Our final example, the strong responder spheroid, is displayed in Figure \ref{fig:credintprog18}.  Here we see how the algorithm treats a function with a steep gradient and a nearly zero tumor volume for the majority of the treatment period.  The scan collection chosen contains days 16, 29, 30, 31, 32, 33, 36, 37, 38, 39, 40, 41, 43, 44, 54; here we show only the first five scans and the final calibration result, as it is clear that the trajectory has stabilized and the uncertainty has been fully reduced after only the first few scans, such that the algorithm can be terminated.  In fact, by including only two scans from the treatment period (days 16 and 29), the algorithm is essentially finished, with respect to both error and uncertainty.

\begin{figure}[htbp]
         \includegraphics[width=0.47\linewidth]{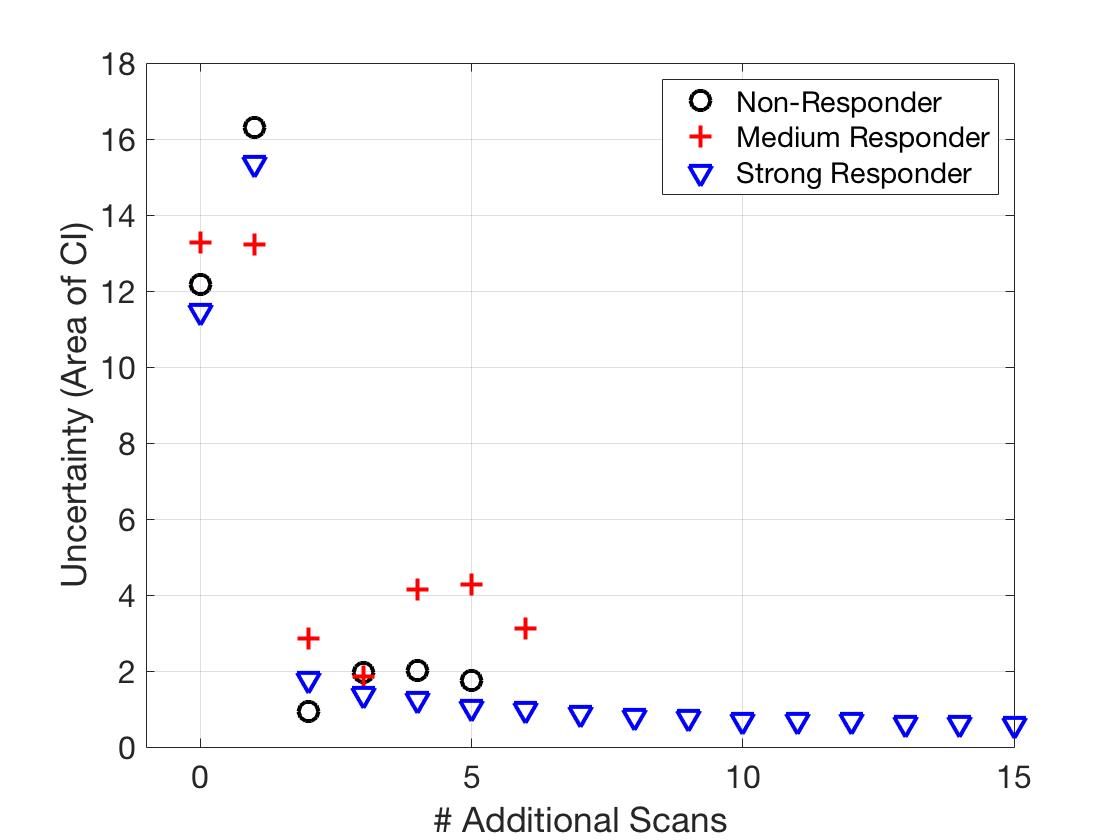}
         \includegraphics[width=0.47\linewidth]{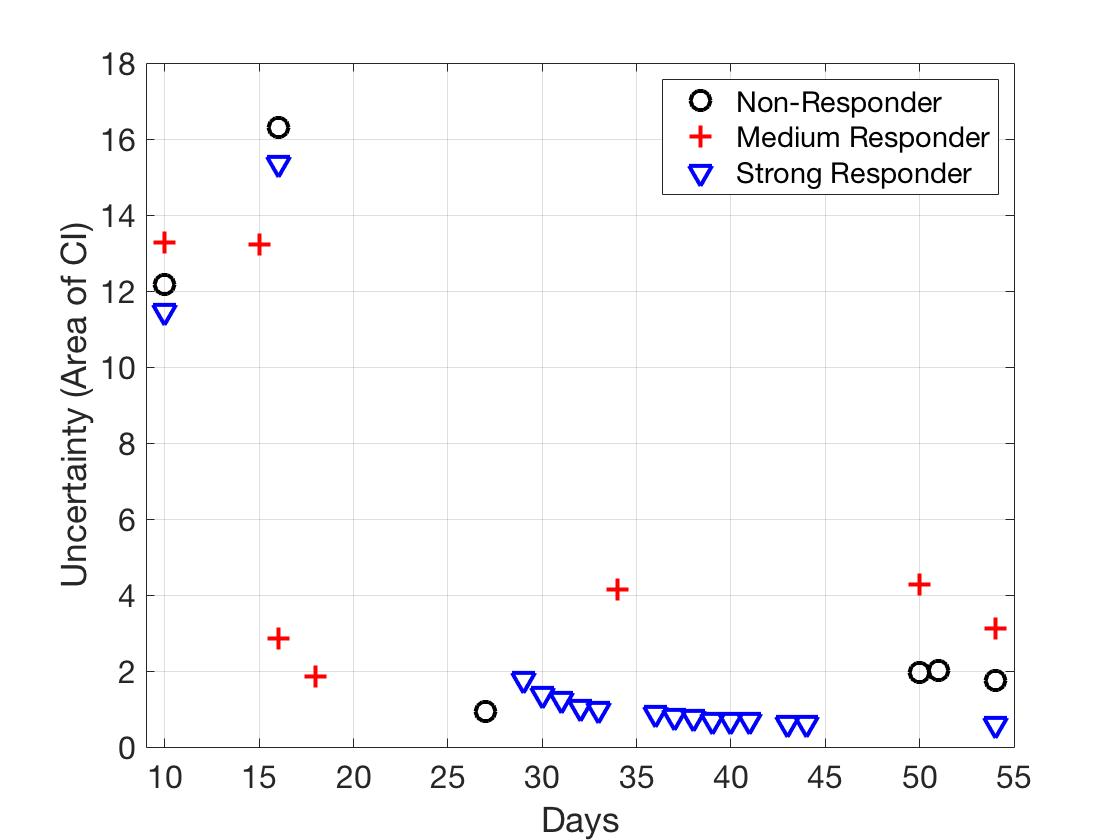}\\
         \includegraphics[width=0.47\linewidth]{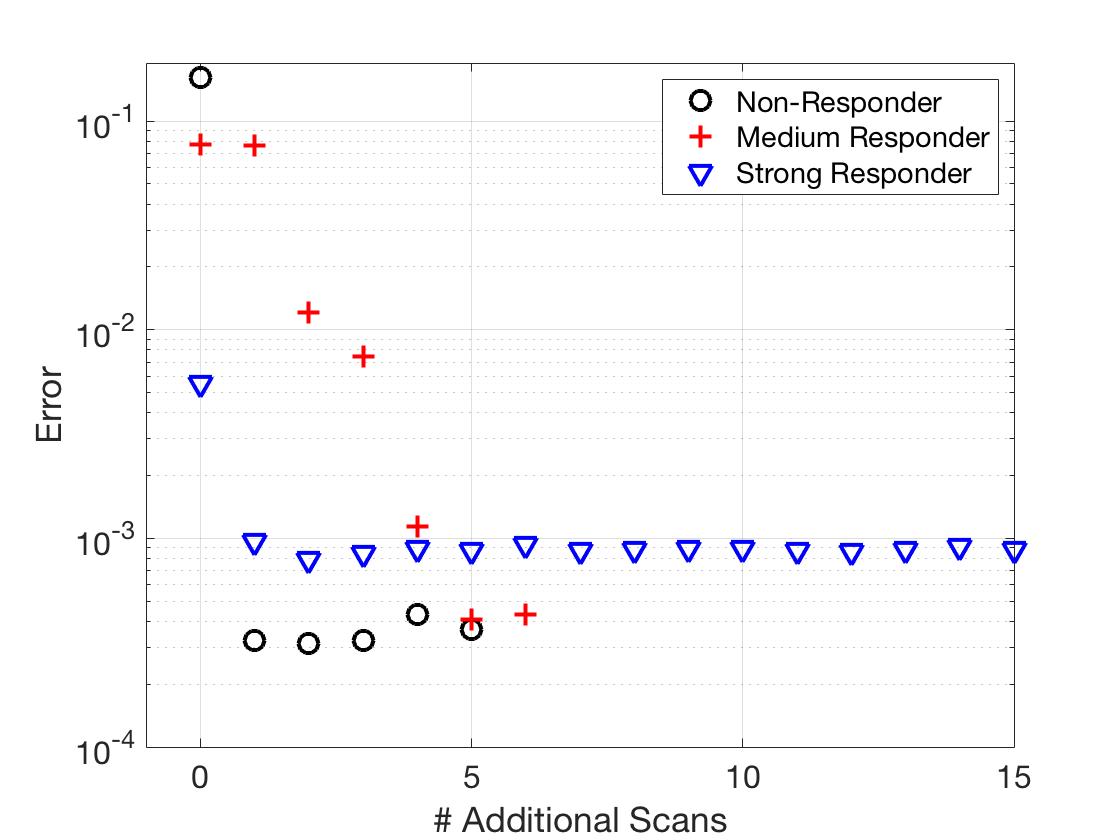}
         \includegraphics[width=0.47\linewidth]{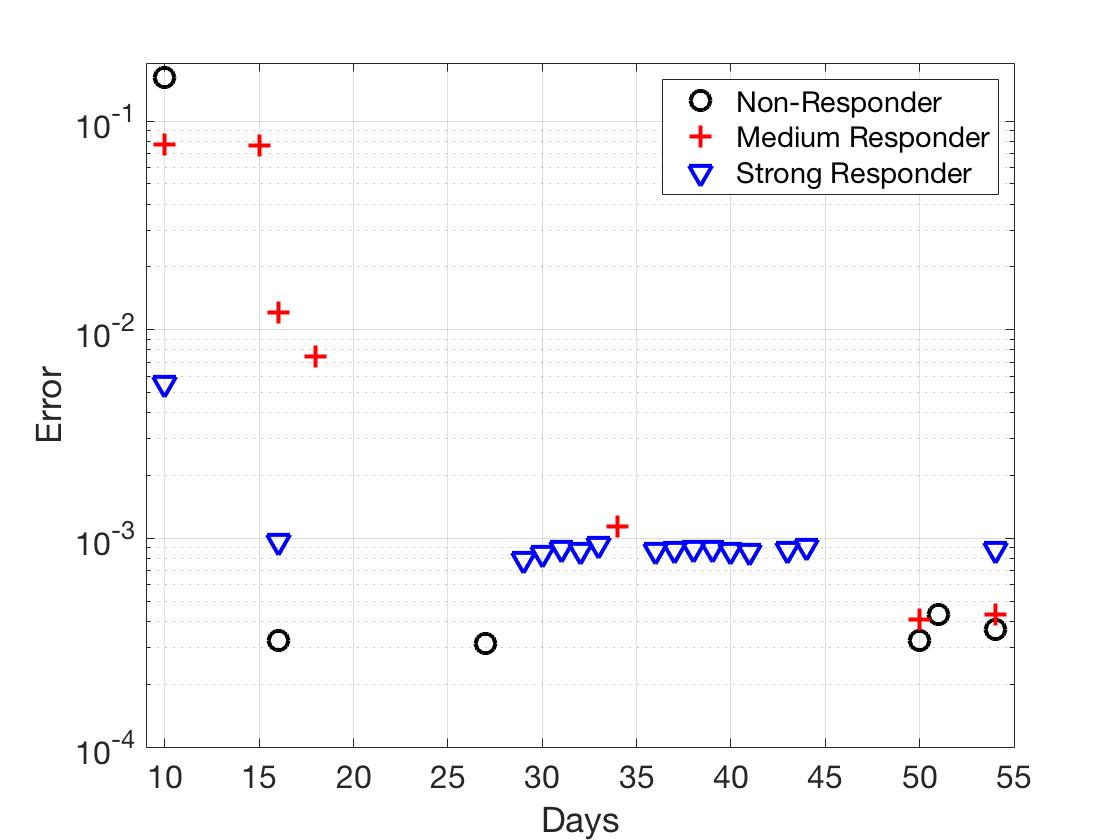}\\
    \caption{Comparing uncertainties and errors over scan progression for all patients with respect to number of scans (left) and day (right). }
    \label{fig:uncertaintycomp}
\end{figure}

Figure \ref{fig:uncertaintycomp} illustrates our uncertainty and error metrics from Section \ref{sec:assessment} for each of the three spheroids. Looking at the metrics plotted with respect to the number of scans (left), the trend becomes clear for both error and uncertainty; both metrics are large for the first two added scans, but decrease greatly with the inclusion of the third scan.  Additionally, we plot each metric with respect to the days (right), so as to better illustrate how early in the treatment period our model can be considered to be near its final state of calibration; recall that our secondary goal is to use not only as few scans as possible, but to finish the calibration as early in the treatment period as possible, to allow for revisions to be made to the treatment regimen, as necessary.  

\begin{figure}[htbp]
         \includegraphics[width=\linewidth]{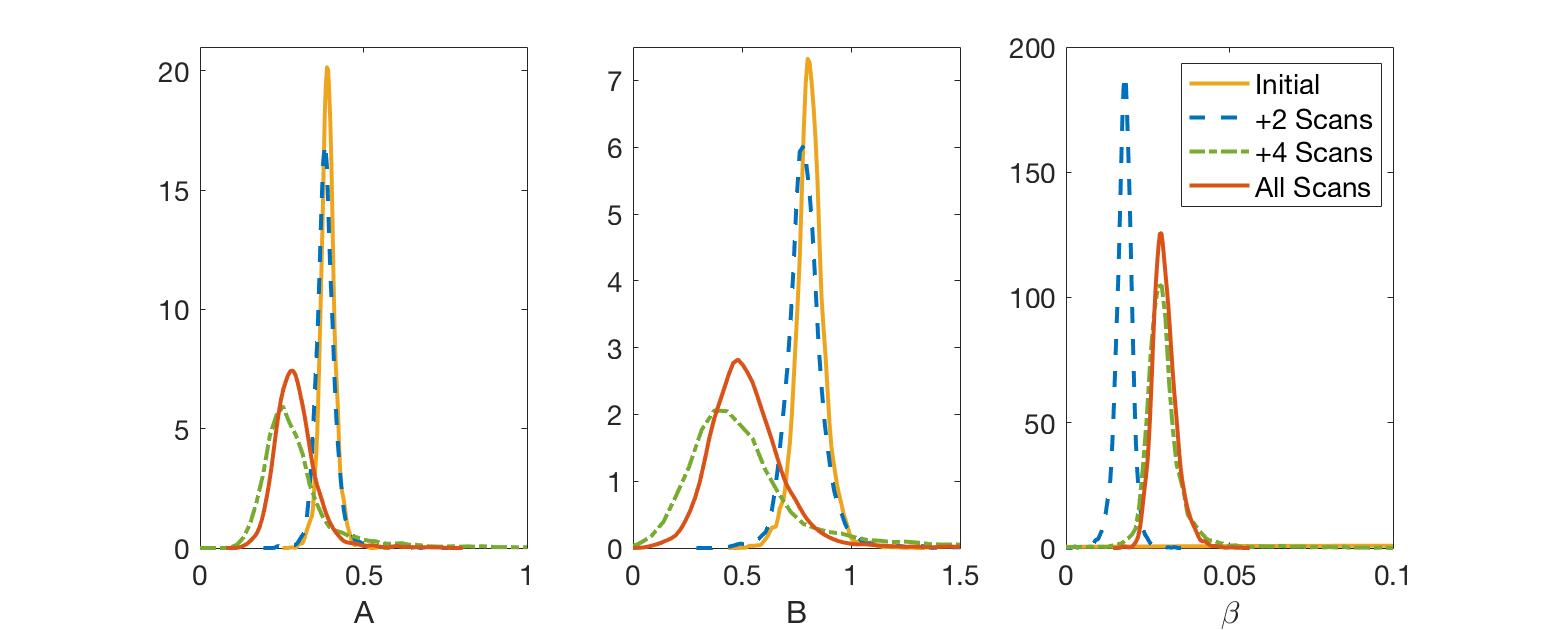}\\
    \caption{Comparing the parameter posterior distributions for the medium responder spheroid as scans are added and the model is re-calibrated. Notice that the initial distribution for $\beta$ is entirely uninformative, as no information has yet been supplied about the regime on which $\beta$ is active. }
    \label{fig:posteriorcomparison}
\end{figure}

To illustrate how the parameter posterior distributions approach convergence as we continue to add scans, we include Figure \ref{fig:posteriorcomparison}, which shows the parameter posterior distributions for four steps of the process for the medium responder spheroid: at the initial calibration, after two added scans, after four added scans, and after all six scans have been included. We note that in the initial calibration, the $\beta$ posterior is entirely non-informative, since we have not yet supplied any information from the time regime on which $\beta$ is active. For all three parameters, we can see how the posteriors shift as additional information is gained. Based on the three-fold goals of (a) reducing uncertainty, (b) reducing error, and (c) waiting for the trajectory to stabilize, it seems reasonable to terminate the algorithm after four scans for the medium responder; this is further supported by observing how the posterior distributions are nearly in alignment for the ``+4 Scans" and ``All Scans" cases in Figure \ref{fig:posteriorcomparison}.  We note that the non-responder and strong responder cases also demonstrate convergence of the posterior distributions as in Figure \ref{fig:posteriorcomparison}, but do so even faster than the medium responder in alignment with their quicker declines in error and uncertainty metrics. 

\subsection{Noise Analysis} \label{sec:noise}

We finish with a brief analysis of how our uncertainty and error metrics are affected by the addition of measurement noise.  For this investigation, we randomly generate the noise from a uniform distribution of varying width, though certainly other noise models such as Gaussian noise could be used. For each data point, $y_{\text{exact}}(t)$, we adjust to create our noisy data point, $$y_{\text{noise}}(t) = y_{\text{exact}}(t)(1+\varepsilon),$$ where $\varepsilon \sim \mathcal{U}(-x,x)$ if we desire $100x\%$ noise.

\begin{figure}[htbp]
\centerline{  
         \includegraphics[width=0.25\linewidth]{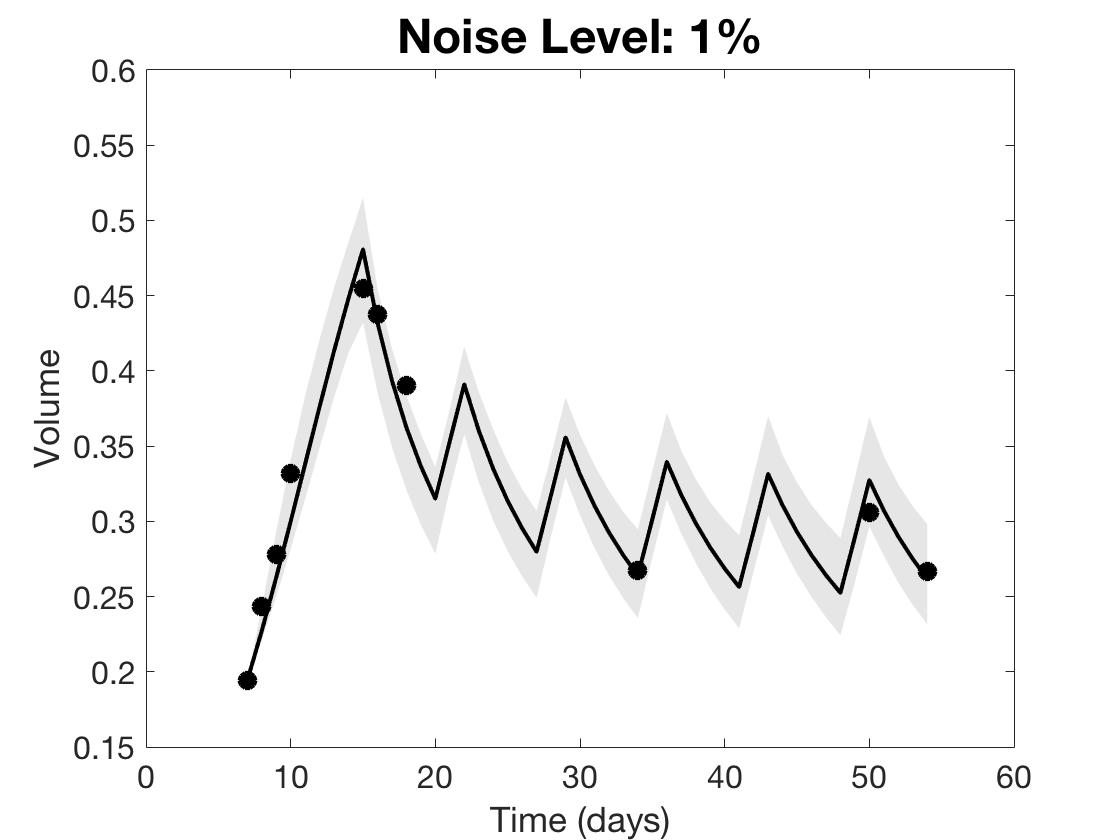}
         \includegraphics[width=0.25\linewidth]{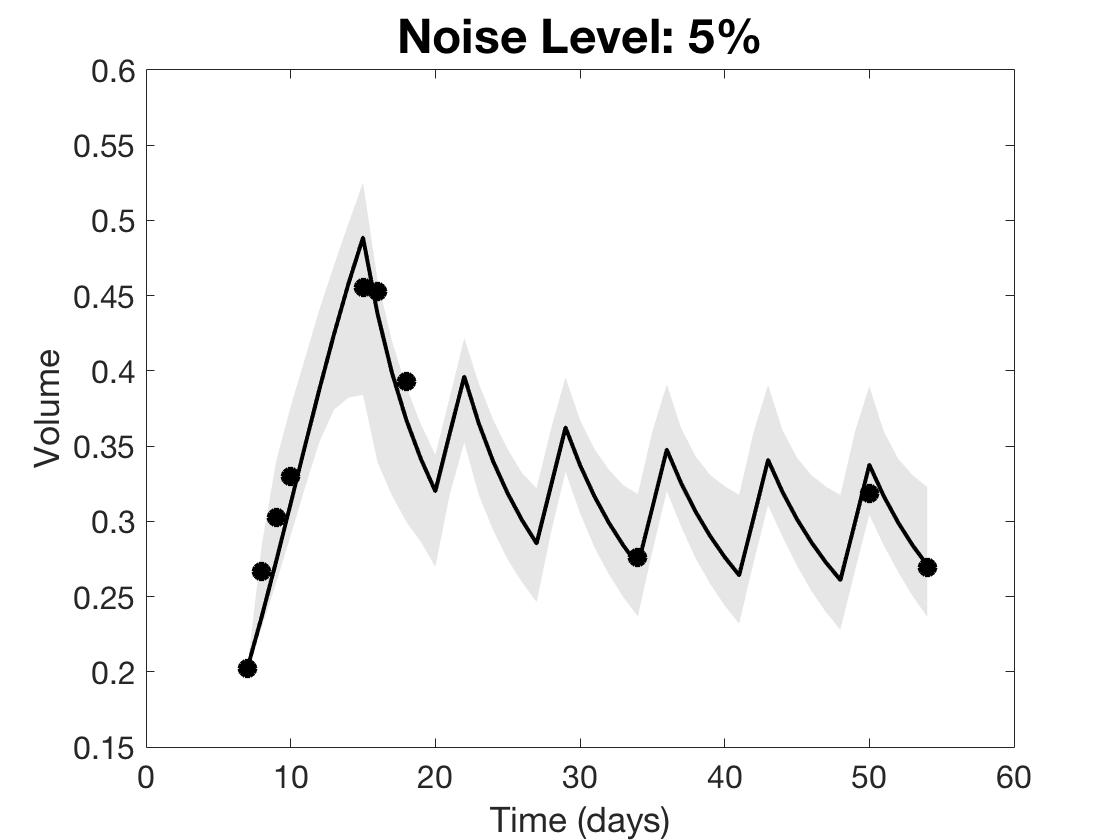}
         \includegraphics[width=0.25\linewidth]{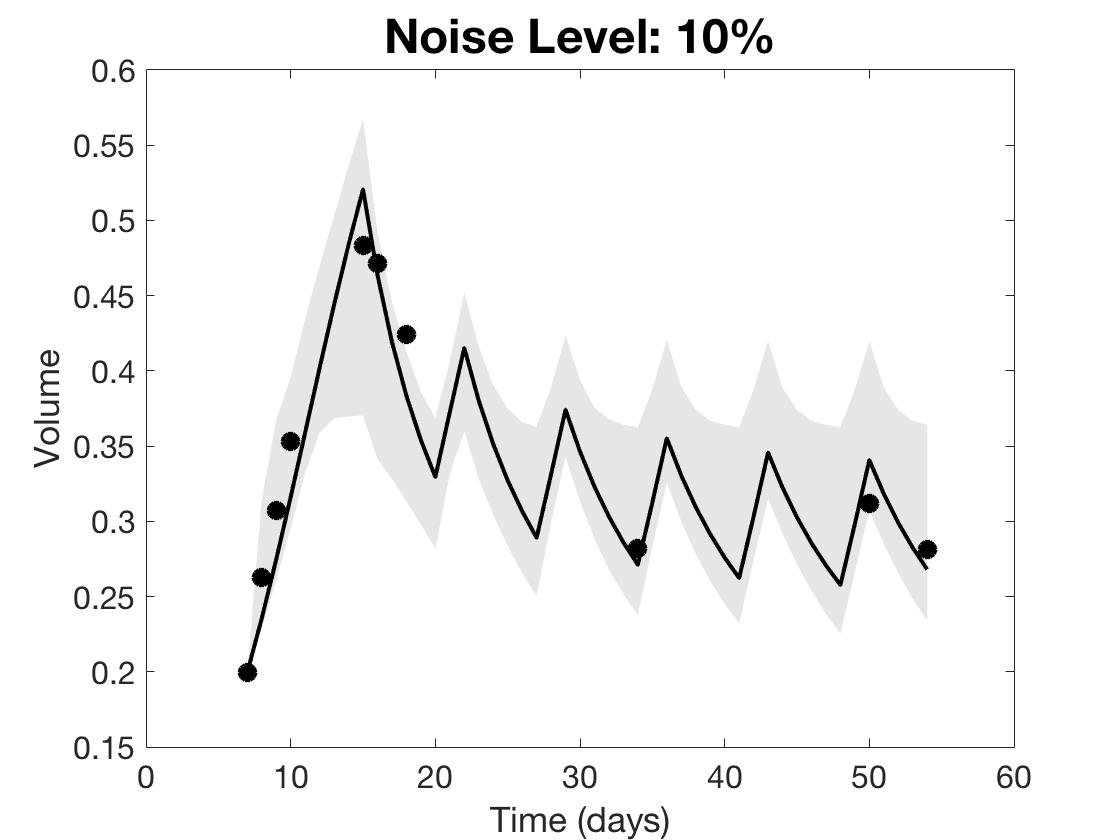}
         \includegraphics[width=0.25\linewidth]{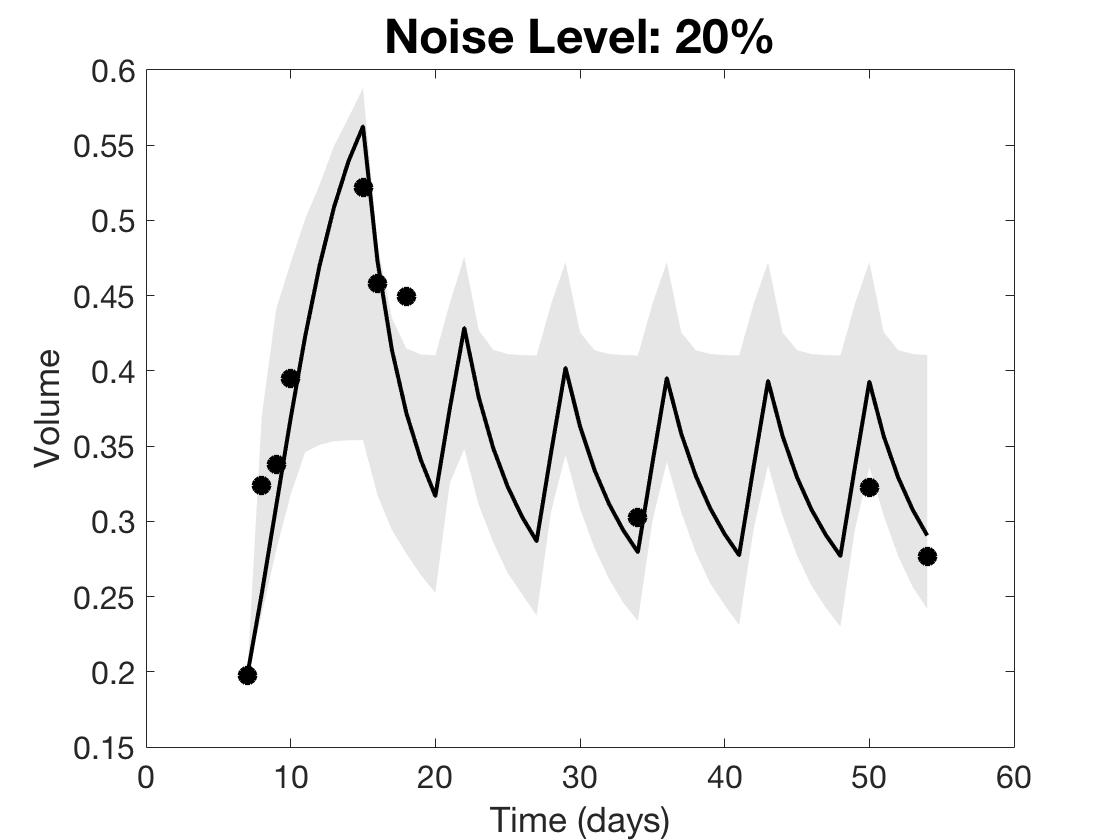}
         }
    \caption{Comparing credible interval widths for sample simulations as data noise level is increased for the medium responder.}
    \label{fig:noisecomp}
\end{figure}

We present the results for our medium responder here, and note that the analysis for the other two examples yields similar results.  For each noise level---1\%, 5\%, 10\%, and 20\%---we run twenty simulations, tracking the area of the credible interval and error in the model fit for each added scan using the scans selected for the medium responder in Section \ref{sec:scorecomp}. Figure \ref{fig:noisecomp} displays a sample simulation from the final model fit for each of the four noise levels. One can observe that both the area of the credible interval and the error between the model fit and the given data increases as the noise level is increased.

To present an overall trend in our metrics as a function of noise level, we report the average area over our twenty simulations of the credible interval at each added scan in Table \ref{tab:noise_unccomp}, and the average error over the twenty simulations at each added scan in Table \ref{tab:noise_errcomp}.  In both cases, we see the general trend of reduction in the metric as the number of scans is increased, as well as an increase in the metric as the noise level grows.  

\begin{table}[!htb]
    \centering
    \begin{tabular}{|c|c|c|c|c|c|c|c|}
    \hline
        \textbf{Ave. Uncertainty} & \textbf{Initial} & \textbf{+1} & \textbf{+2} & \textbf{+3} & \textbf{+4} & \textbf{+5} & \textbf{+6} \\
        \hline 
        \textbf{1\% Noise} & 17.04 & 16.43 & 2.77 &  1.82 &	4.52 &	4.63 & 3.39 \\
        \hline
        \textbf{5\% Noise} & 18.37 & 16.75 & 4.38 &	2.43 &	4.56 &	4.72 & 3.43 \\
        \hline
        \textbf{10\% Noise} & 16.68 & 17.52 & 5.36 & 3.44 & 4.72 & 4.79 & 3.94\\
        \hline 
        \textbf{20\% Noise} & 17.13 &	17.95 &	12.18 &	5.73 &	5.54 &	5.83 & 	5.03 \\
        \hline
    \end{tabular}
    \caption{Measuring average uncertainty over 20 simulations using total area of credible interval as a metric for various levels of noise. (Medium responder.) }
    \label{tab:noise_unccomp}
\end{table}

\begin{table}[!htb]
    \centering
    \begin{tabular}{|c|c|c|c|c|c|c|c|}
    \hline
        \textbf{Ave. Error} & \textbf{Initial} & \textbf{+1} & \textbf{+2} & \textbf{+3} & \textbf{+4} & \textbf{+5} & \textbf{+6} \\
        \hline 
        \textbf{1\% Noise} & 0.0692 &	0.0636 &	0.0123 & 	0.0074 &	0.0012 &	0.0004 &	0.0004\\
        \hline
        \textbf{5\% Noise} & 0.0685 & 	0.0671 &	0.0116 &	0.0072 & 	0.0012 & 	0.0005 &	0.0005  \\
        \hline
        \textbf{10\% Noise} & 0.0673 & 	0.0766 &	0.0125 & 0.0072 &	0.0014 &	0.0007 &	0.0007 \\
        \hline 
        \textbf{20\% Noise} & 0.1131 &	0.0928 &	0.0267 &	0.0084 &	0.0027 &	0.0016 &	0.0015 \\
        \hline
    \end{tabular}
    \caption{Measuring average error over 20 simulations using mean squared error for all days 7-54 as a metric for various levels of noise. (Medium responder.)}
    \label{tab:noise_errcomp}
\end{table}

While we have no issues fitting the model to the data in any of the noise levels tested (and all final parameter estimates are verified to be practically identifiable), we note that an increase in the uncertainty of the model trajectory may lead to less meaningful results in terms of the predictive power of the model.  Thus, we urge future investigators to be cognizant of the level of measurement noise in their data, and to consider how the uncertainty and error might be affected by the level of refinement one is able to achieve when collecting data.

\section{Discussion} \label{sec:discussion}

In this work, we propose a gradient-based score function to determine an optimal scanning protocol for cancer patients treated with radiotherapy. This score function is used within a Bayesian sequential design framework, to choose tumor scanning days that maximize the mutual information provided by the scan, while simultaneously incorporating a penalty for scans obtained late in the treatment period. We test this methodology by generating a wide range of synthetic tumor spheroid data from a detailed CA model and then sequentially calibrating a simple ODE model using this data. 

Using both error and uncertainty to assess the predictive power of the model, we show that our algorithm chooses scans that can be used to calibrate the low-fidelity model accurately and efficiently.  For weak and strong responders to treatment, both the error and uncertainty shrink significantly once two additional treatment scans have been included in the calibration, highlighting the efficiency of the algorithm. Our results reinforce the accuracy of the calibration by showing that the parameter values stabilize at two scans, with the inclusion of additional scans leading to minimal changes in the parameter values. For the medium responder, four additional scans are needed to achieve this same level of certainty and accuracy, since the dynamics have not settled as quickly as the weak and strong counterparts.

The gradient-based score function presented in this work improves upon a previous score function proposed in \cite{ChoJCM}. 
Since the optimal value of $k$ seemed to be closely related to the shape of the data, we hypothesized that this parameter value might be eliminated by gathering information about the gradient of the data along the way. That is, if we can approximate the steepness in the data (i.e., the strength of the response to radiotherapy), we can use that approximation to adjust our score function to select data points close to the current time step (which might be favored in scenarios where the gradient is steep and data is informative) or to select points located further out in time (which could be preferred if the gradient is relatively flat and we don't expect to see much change in the next few days). The incorporation of this information to the score function not only allows us to eliminate the penalization parameter $k$, but also aligns well with our goal of reducing unnecessary data collection in non-informative regions.

In addition to updating the score function to reflect information about the gradient, we have also provided a more robust and thorough verification of the algorithm in this investigation.  The error-based verification metrics of the previous study are now supported by an uncertainty-based analysis, which relies on the propagation of parameter posterior distributions through the model to produce a 95\% credible interval of model trajectories. The forward-looking nature of uncertainty analysis provides a more practical means of deciding when the algorithm might be terminated. Additionally, we analyze how the uncertainty in the model is affected by the addition of measurement noise.  By considering the combination of model error, model uncertainty, and the stabilizing of model trajectories, we illustrate how these metrics might be used to decide when enough data has been collected to suit the purposes of the investigator.

In this work, we have applied our methodology to synthetic tumor spheroid datasets, generated from a CA model, and we are encouraged by the accuracy and efficiency of the algorithm and the improvements made with the gradient-based score function. In future work, we plan to test the robustness of this methodology on experimental and clinical tumor data. If our results in an idealized setting translate to noisy, real-world data, this calibration framework may eventually be used in the clinic to inform tumor scanning protocols for patients receiving radiotherapy. We anticipate that this will improve the efficiency of clinical data collection and will help to make more accurate predictions about individual patient response to radiotherapy. Additionally, we plan to test this methodology on other low-fidelity models that incorporate additional tumor characteristics and/or different treatment modalities. This will increase the flexibility of our methodology, enabling its application to many different settings in clinical oncology.  
\backmatter














\begin{appendices}

\section{Parameter Values for Agent-Based Model Data Generation} \label{app:params}

\begin{table}[!htb]
\begin{center}
\begin{tabular}[c]{|c|c|c|c|}
        \hline
        \textbf{Parameter} &\textbf{Description} & \textbf{Value} & \textbf{Units} \\

\hline 
 $l$ & Cell size & 0.0018 & cm \\
\hline
 $L$ & Domain length & 0.36 & cm \\
\hline
 $\bar{\tau}_{cycle}$& Mean cell cycle time   & Varies & h \\ 
\hline 
 $c_{\infty}$& Background O$_2$ concentration & $2.8\times 10^{-7}$ & mol cm$^{-3}$  \\
\hline 
 $D$ & O$_2$ diffusion constant & $1.8\times 10^{-5}$ & cm$^2$s$^{-1}$\\
\hline 
 $c_Q$ & O$_2$ concentration threshold for proliferating cells & 1.82$\times10^{-7}$ & mol cm$^{-3}$ \\
\hline 
$c_N$ & O$_2$ concentration threshold for quiescent cells & 1.68$\times10^{-7}$ & mol cm$^{-3}$ \\
\hline 
 $\kappa_P$ & O$_2$ consumption rate of proliferating cells & 1.0$\times10^{-8}$ &mol cm$^{-3}$s$^{-1}$\\
\hline 
 $\kappa_Q$ & O$_2$ consumption rate of quiescent cells & 5.0$\times10^{-9}$ & mol cm$^{-3}$s$^{-1}$  \\
\hline 
 $p_{NR}$ & Rate of lysis of necrotic cells & 0.01  & hr$^{-1}$ \\
\hline 
\end{tabular}
\caption{A summary of the parameters used in the CA Model and their default values. Parameter values are estimated using experimental data from the prostate cancer cell line, PC3, in \cite{Paczkowski2021}.}
\label{table:CA_pars}
\end{center}
\end{table}




\end{appendices}


\bibliography{sn-article.bib}


\end{document}